\documentclass[%
 reprint,
 amsmath,amssymb,
 aps,
]{revtex4-2}
\usepackage{natbib}
\usepackage{graphicx}
\usepackage{dcolumn}
\usepackage{mathtools}
\usepackage{xcolor}
\usepackage{bm}

\begin{document}


\title{Mutual Consistency of Multiple Visual Feature Maps Constrains Combined Map Topology}

\author{X. Liu}
\email{xliu9362@uni.sydney.edu.au}
\affiliation{School of Physics, The University of Sydney, NSW 2006, Australia}
\affiliation{Center for Integrative Brain Function, The University of Sydney, NSW 2006, Australia}
\email{peter.robinson@sydney.edu.au}
\author{P. A. Robinson}
\affiliation{School of Physics, The University of Sydney}
\affiliation{Center for Integrative Brain Function, The University of Sydney, NSW 2006, Australia}

\date{\today}

\begin{abstract}
The topologies permitted in joint ocular dominance (OD), orientation preference (OP), and direction preference (DP) maps in the primary visual cortex (V1) are considered, with the aim of finding a maximally symmetric periodic case that can serve as the basis for perturbations toward natural realizations. It is shown that mutual consistency of the maps selects just two possible such lattice structures, and that one of these is much closer to experiment than the other. This comprises a hexagonal lattice of alternating positive and negative OP singularities, with each unit cell or hypercolumn containing four such singularities, each radiating three DP discontinuities that follow OP contours and end at OP singularities of opposite sign. Other DP discontinuities emanate at 90 degrees to the midpoints of the ones that link OP singularities, and cross OP contours perpendicularly. These features explain experimentally observed relationships between DP discontinuities and OP contours, including sudden approximately 90-degree changes of direction in the former.
\end{abstract}

\maketitle

\section{\label{introduction}Introduction}
Visual stimuli are mapped in a one-to-one manner to primary visual cortex (V1) via the retinotopic map, which preserves topology but disproportionately represents the center of the visual field relative to the periphery. V1 processes multiple stimulus features, including ocular dominance (OD; i.e., preference for stimuli from a particular eye), edge orientation, direction of motion, spatial frequency, and color. This processing is arranged in such a way that all features from each segment of the field of view are processed in a small neighborhood in V1, termed a hypercolumn \citep{Gilbert_patchy_connectn,Hubel_column_arrangm_1962,hubel_sequence_1974,Miikkulainen_visual_maps}. Each hypercolumn is of order 1 -- 2 mm in linear size and contains cells sensitive to each possible feature; hypercolumns do not have structural boundaries, but are analogous to the unit cell of a crystal lattice, being defined only in that each contains cells with a full set of stimulus responses in aggregate.  Indeed, previous studies have suggested that V1 can be approximated by a lattice of hypercolumns \citep{bressloff_visual_2002,bressloff_functional_2003,Veltz_2015}. 

Cells that respond preferentially to a given feature are arranged in a column spanning the cortical thickness and selectivity to specific features changes in an orderly way within each hypercolumn, and across V1 more generally, to form a feature map. Each cell typically responds to several features via these overlapping maps, which are described in detail in the next section. 

Because each hypercolumn must contain cells that respond to all possibilities for each of multiple features, the various feature maps are not arbitrary, but are closely interrelated. For example, motion of an oriented edge is only detectable if a component of the motion is perpendicular to the edge \cite{kisvarday2001calculatingDP,swindale2003spatialpattern,weliky1996DPsystematic}, so maps of orientation preference (OP) and direction preference (DP) are inextricably linked, and both must be consistent with the ocular dominance (OD) map. We must thus ask what topologies of the combined maps are allowed under the constraints of mutual consistency, complete feature coverage in each hypercolumn, and periodicity. We seek exact periodicity in order to provide a starting point for perturbation analyses via which topological and geometric lattice imperfections can be introduced. Although previous studies \citep{bressloff2001geometric,bressloff_PhysRevLet_2002,bressloff_visual_2002, bressloff2003spatially,bressloff_functional_2003,Veltz_2015} have employed regular planar lattices to investigate the spontaneous formation of activity patterns in V1, these studies mostly focused on modeling the combined effects of local and long-range lateral connections within and between hypercolumns, as well as the impact of lattice symmetry on the possible activity patterns. These authors studied the lattice structures of cytochrome oxidase (CO) blobs, OP, and spatial frequency (SF) preference, but not DP, and they did not explore the mutual consistency of feature maps. The present study is complementary in that it investigates how mutual consistency of OD, OP, and DP maps constrains their characteristics and possible regular lattice structures.

In this paper, we first unify different experimental observations of the feature maps, and discuss its main characteristics in Sec.~\ref{sec:maps}. Then, in Sec.~\ref{sec:topology} we consider how mutual consistency of the OD, OP, and DP maps restricts the possible structure of the combined multi-feature map. Moreover, we explore how imposition of symmetry further constrains the combined map, ultimately leading to a unique topology. We then compare the results with experiments and discuss the effects of lattice imperfections introduced when we relax some idealizations. Finally, the main findings are summarized and discussed in Sec.~\ref{conclusions}. Throughout this study, we focus only on simple cells in V1, which respond best to oriented bars and have separated antagonistic excitatory and inhibitory subregions in its receptive field \citep{hubel1962receptive,hubel1968receptive}.

\section{Feature Maps}
\label{sec:maps}
In this section we briefly summarize key experimental findings relating to observed OD, OP, and DP maps and their relationships. We are not concerned with the structure of feature-preference columns vs.~depth through cortical layers, but only with the dependence of feature preference on location on the surface of V1.

OD maps reflect the arrangement of neurons with particular left- or right-eye preference. In \cite{hubel1968receptive} it was first found that neurons that respond to same eye preference (left, L, or right, R) are aggregated into columns that span several cortical layers vertically, and are laterally arranged as approximately parallel alternating stripes, in a pattern that is reminiscent of a fingerprint. The width of each (L or R) OD stripe is approximately 0.4 mm in macaque monkeys and 0.8 -- 1.2 mm in humans \citep{adams_complete_2007,cheng2001human_Od,Horton_OD,horton_intrinsic_1996,Hubel_column_arrangm_1962,hubel1974uniformity,hubel_ferrier_1977}, giving overall spatial periodicities of twice that distance perpendicular to the stripes between points of the same OD preference. OD selectivity is highest in the middle of each OD stripe, and decreases towards the stripe borders where binocular cells are found that respond to both eyes \citep{crair1997ocular,shatz1978ocular}. This enables the OD map to be continuous by making the net OD preference zero at boundaries where there would otherwise be a discontinuity.

OP columns consist of neurons that respond to particular edge orientations of features in the visual field. OP normally varies continuously as a function of the cortical position and covers the complete range of  orientations from $0^\circ$ to $180^\circ$ within each hypercolumn  \citep{hubel_sequence_1974,hubel_ferrier_1977,obermayer_geometry_1993}. Optical imaging \citep{blasdel_orientation_1992,bonhoeffer_iso-orientation_1991,bonhoeffer_layout_1993, swindale_review_1996} shows that key features of the OP map are that: (i) OP varies azimuthally around singularities, often termed pinwheel centers (PWCs), where all preferences meet; (ii) the OP in each pinwheel increases either clockwise (negative PWC) or counterclockwise (positive PWC) and around 85\% of neighboring PWCs have opposite signs \citep{gotz_d-blob_1987,Gotz_1988,blasdel_orientation_1992}; (iii) neighboring pinwheels are usually directly connected by iso-orientation contours that are relatively straight; (iv) there is low OP selectivity at PWCs near which all OPs occur, consistent with achieving map continuity by means of selectivity going to zero at the PWCs;  mechanistically, this could be the result of averaging the responses of nearby neurons with a wide range of OPs \citep{maldonado1997orientation,ohki2006highlyorderpinwheel}; and (v) occasionally the smooth OP map is interrupted by lines, termed OP fractures, perpendicular to which OP changes of more than $45^\circ$ occur over short distances \citep{blasdel1986voltage,blasdel_orientation_1992}. However, later studies \citep{shmuel1996functional, weliky1996DPsystematic} found that most OP discontinuities are at PWCs and fractures are much less common. It is thus possible that a significant proportion of regions identified as fractures actually contain rapid, but smooth, changes of OP.
 
Certain V1 neurons have been found to be selective for the direction of motion of edges in the visual field in cats, ferrets, and humans \citep{shmuel1996functional,swindale1987surface,wang2014motion, weliky1996DPsystematic}. The resulting DP map is thus closely tied to the OP map and thus varies smoothly over most of V1. A key difference is that two possible directions of motion can correspond to one edge orientiation, so a $360^\circ$ range of DP must be accommodated within each hypercolumn. This results in the widespread presence of lines, termed DP fractures, along which DP suddenly jumps by $180^\circ$; these fractures usually radiate from PWCs, and often connect them, but can also terminate at other locations \citep{shmuel1996functional,swindale1987surface,wang2014motion, weliky1996DPsystematic}. 

OD, OP, and DP maps are closely interrelated because they must be mutually consistent and must represent a full range of features in each hypercolumn. Experimental observations of relationships between OP and OD include that most PWCs lie near the middle of OD stripes and lines of constant OP usually cross OD boundaries nearly perpendicularly \citep{bartfeld_relationships_1992,blasdel_orientation_1992,obermayer_geometry_1993}. Both these aspects are seen in Fig.~1(A), which shows the combined OD-OP map in adult macaque monkeys cortex. 

\begin{figure}
\includegraphics[width=0.4\textwidth]{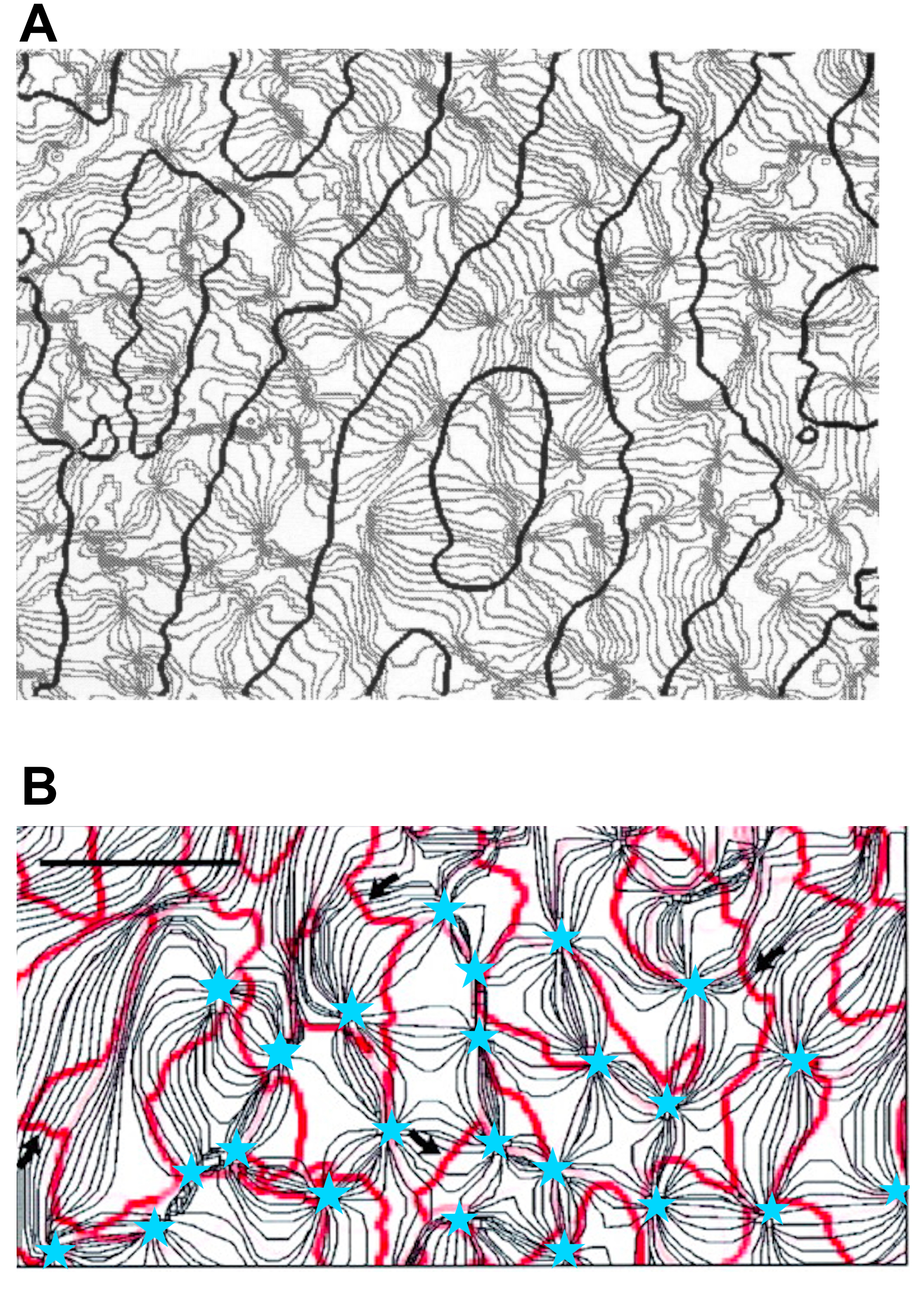}
\caption{\label{fig:experim_op_dp} Experimental feature maps in V1. (A) Combined OD-OP map from \cite{obermayer_geometry_1993}. The bold black lines are the OD borders, and the gray curves are iso-orientation contours. (B) Combined OP-DP map of adult cat, adapted with permission from \cite{kisvarday2001calculatingDP}. The OP and PWCs are represented by the thin black contours and the blue stars, respectively. Thick red lines are DP fractures. Some examples of DP fractures that intersect iso-orientation contours at steep angles are marked by black arrows.}
\end{figure}

Figure \ref{fig:experim_op_dp}(B) shows an experimental OP contour map overlaid with DP fractures of an adult cat \citep{kisvarday2001calculatingDP}. The organization of DP map is strongly constrained by the arrangement of the OP map. As noted above, most fractures end near/at PWCs, but some terminate between PWCs. Most of the DP fractures tend to run parallel to iso-orientation contours, and separate subregions with opposite DP, while a small proportion of DP fractures intersect the OP contours at steep angles as indicated by black arrows in Fig.~\ref{fig:experim_op_dp}(B). These DP fractures separate regions with opposite DP \citep{kisvarday2001calculatingDP,shmuel1996functional,swindale1987surface,weliky1996DPsystematic}. 

\begin{figure}
\includegraphics[width=0.4\textwidth]{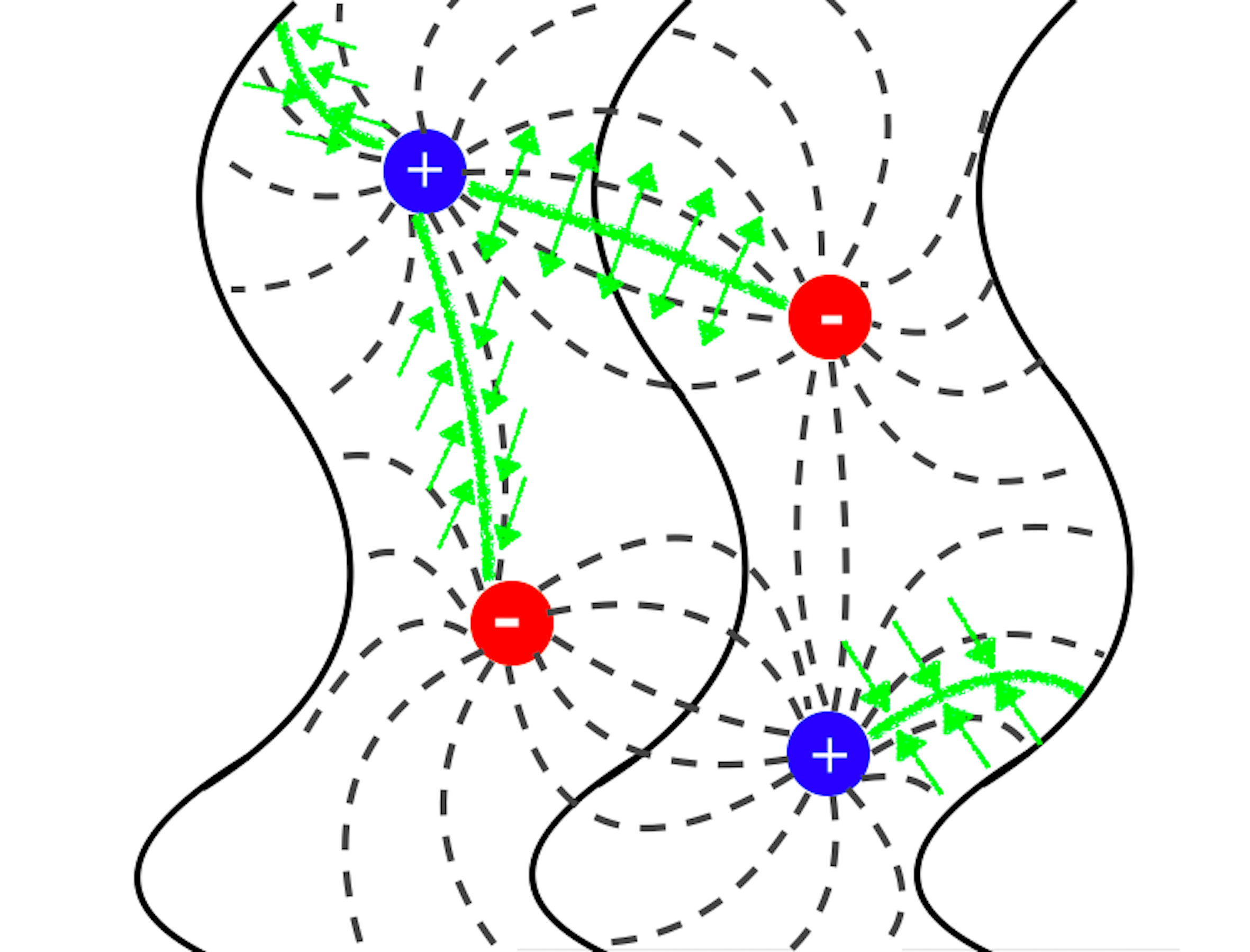}
\caption{\label{fig:schem_op_od_dp} Schematic of the combined OD-OP-DP map in one hypercolumn containing four PWCs. The thin solid black lines represent the borders of the left and right OD stripes, within each of which there are two PWCs with opposite signs, as shown by the heavy dots. Dashed lines represent OP contours, and the thick green contour lines and arrows show DP fractures and adjacent direction preferences.}
\end{figure}

To bring together the experimental observations mentioned above, we show a schematic of a combined OP-OD-DP map within a hypercolumn in Fig~\ref{fig:schem_op_od_dp}. The left and right OD stripes run from top to bottom and within each there are two pinwheels with opposite signs. Dashed lines indicate contours of constant OP and DP directions are indicated by arrows. Each PWC is the origin of one or three DP fractures. 

\section{Topology of Mutually Consistent Multiple Feature Preference Maps}
\label{sec:topology}
Here we progressively build an idealized, mutually consistent, exactly periodic OD-OP-DP map across a lattice of hypercolumns and discuss its properties. At each stage, we discuss connections to experimental observations, including those discussed in Sec.~\ref{sec:maps}.

\subsection{Possible OD-OP Topologies}
To allow for maximal symmetry, we first approximate OD stripes as being parallel and of equal width, with a pair of such stripes in each hypercolumn. Each OD stripe must then contain two PWCs of opposite sign, located close to the midlines of the hypercolumns, each having PWCs of opposite sign as its nearest neighbors both within and outside the hypercolumn. 

\begin{figure}[ht!]
\includegraphics[width=0.25\textwidth]{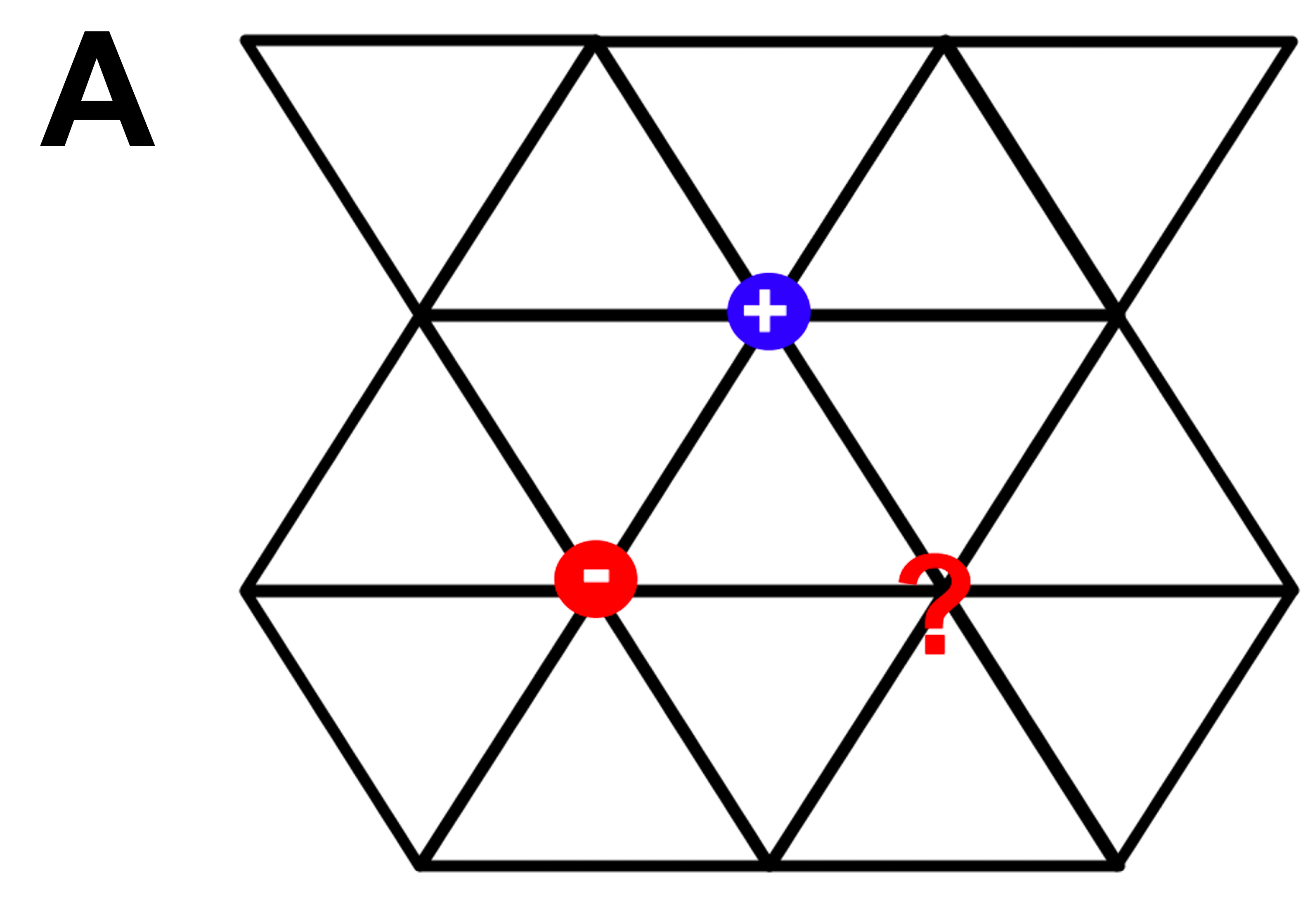}
\includegraphics[width=0.25\textwidth]{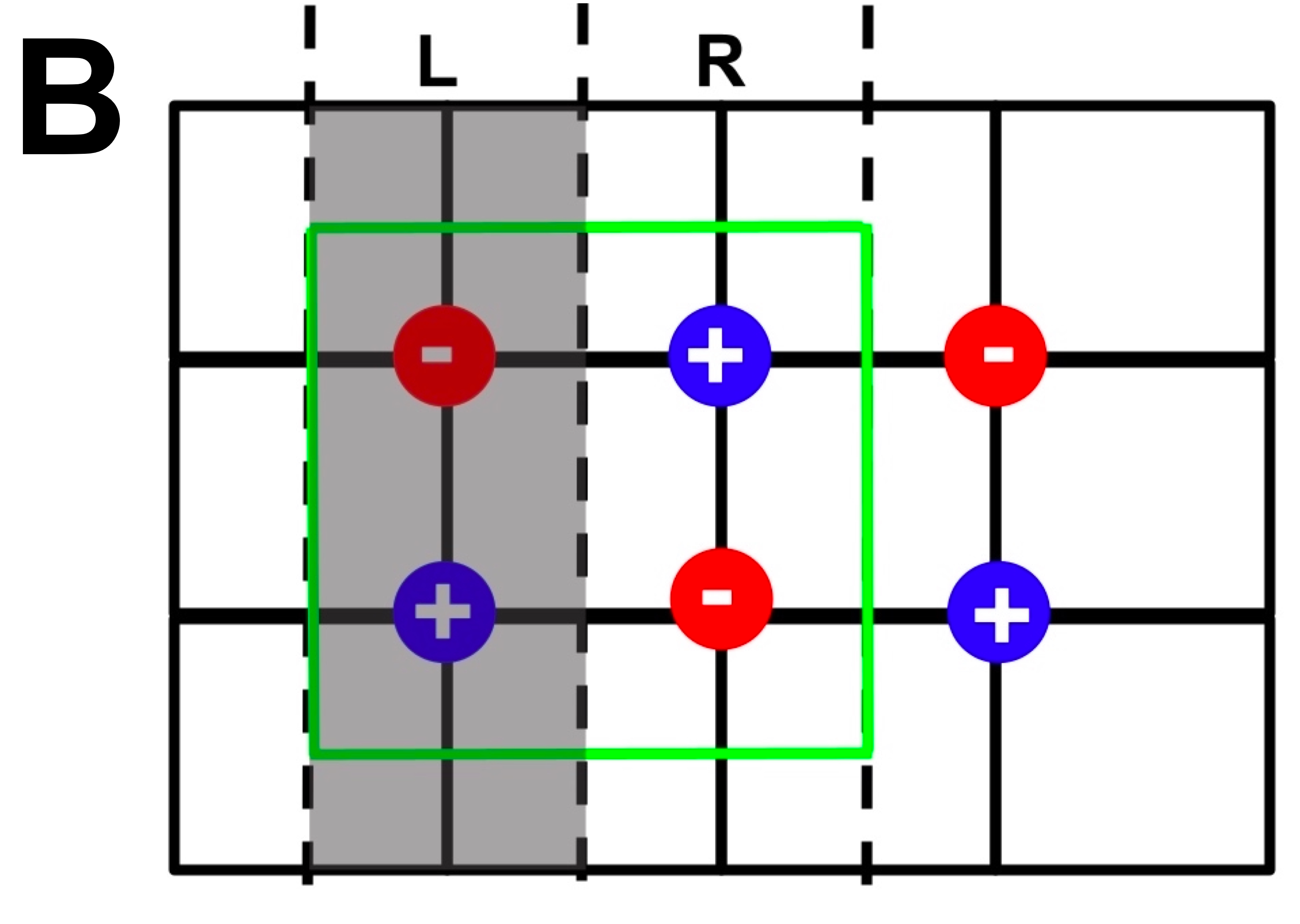}
\includegraphics[width=0.32\textwidth]{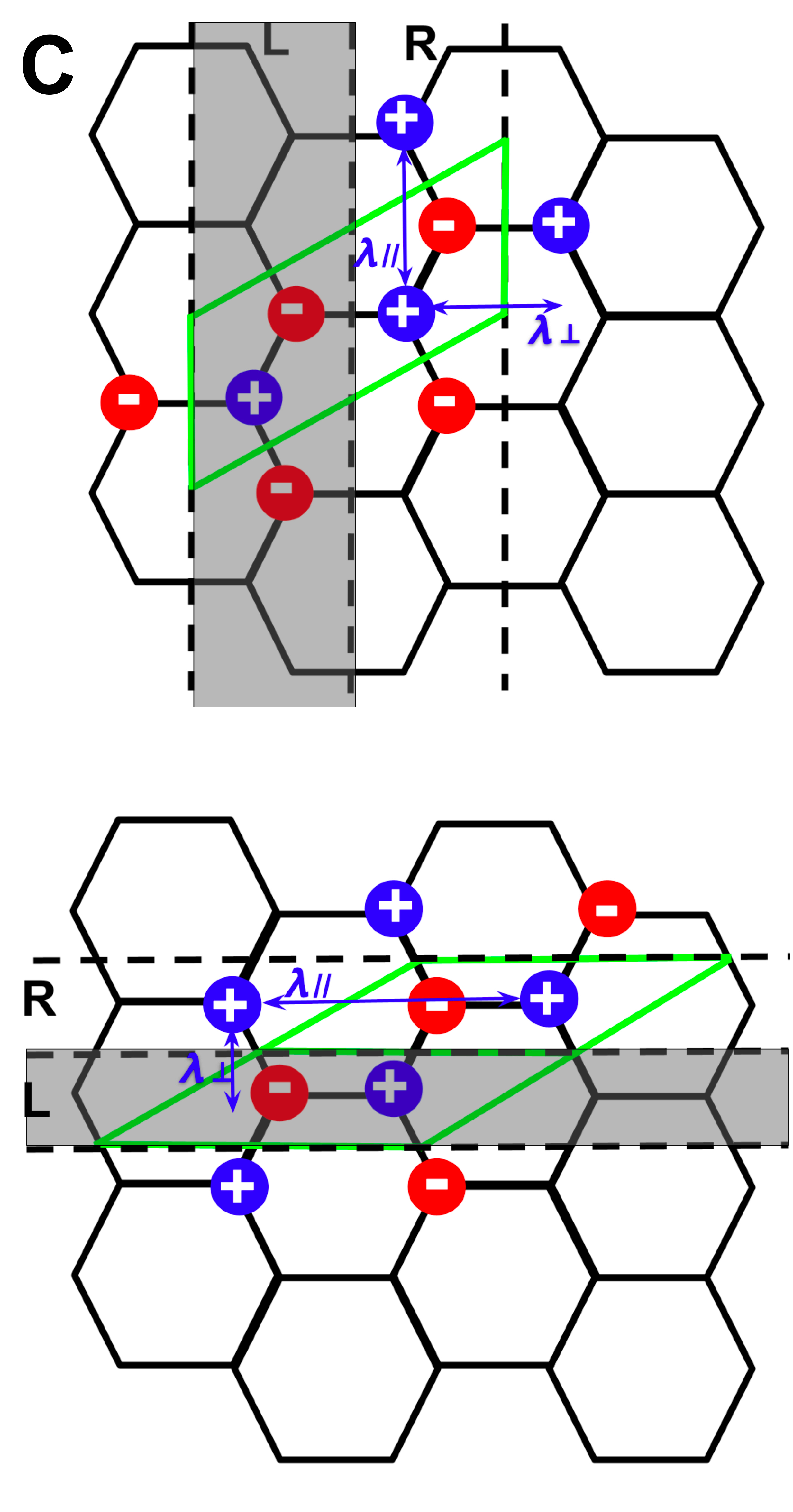}
\caption{\label{fig:hypercol_grid} Schematic of possible hypercolumn topologies formed from lattice of regular polygons. The blue and red dots indicating positive and negative PWCs, respectively. \textbf{(A)} Triangular lattice, showing that it is impossible for all neighboring PWCs to have opposite signs.  \textbf{(B)} Square lattice. The gray and white stripes bounded by dashed lines show a pair of left and right OD stripes. The large green square outlines a hypercolumn containing four pinwheels. \textbf{(C)} Hexagonal lattice showing two possible hypercolumn arrangements as green parallelograms. Blue arrows indicate distances between positive PWCs parallel ($\lambda_{\parallel}$) and perpendicular ($\lambda_{\perp}$) to the OD stripe borders, respectively.}
\end{figure}

The arrangement of hypercolumns into a lattice is restricted by the requirement that PWCs alternate in sign. Only triangular, square, and hexagonal lattices of PWCs are topologically distinct in two dimensions (2D). Of these, the triangular lattice cannot satisfy the need for PWC signs to alternate, as seen in Fig.~\ref{fig:hypercol_grid}(A), a situation that is exactly analogous to frustration in a triangular lattice of antiferromagnetic spins \cite{balents2010spin,wannier1950antiferromagnetism}. Therefore, we eliminate the triangular lattice from further consideration. On the other hand, the square and hexagonal lattices do allow PWC signs to alternate and a number of studies have used them to model various aspects of V1 structure, function, and activity, including during visual hallucinations  \citep{bressloff_visual_2002,bressloff_what_2002,bressloff_functional_2003,Veltz_2015}. 

For the square and hexagonal topologies, we show the possible orientations of OD stripes as gray and white stripes in Fig.~\ref{fig:hypercol_grid}(B) and (C). Aside from rotations and translations that leave the system unchanged, there is only one topologically distinct possibility in the square case, with a hypercolumn outlined in the green square, which contains four pinwheels in Fig.~\ref{fig:hypercol_grid}(B), and OD bands as shaded and labeled. For the hexagonal case there are two topologically distinct OD-OP maps whose corresponding hypercolumns are shown as green parallelograms in Fig.~\ref{fig:hypercol_grid}(C), corresponding to horizontal and vertical OD stripes, respectively. In the horizontal case, PWCs lie exactly on the midline of the OD stripes shown, but these are narrow, each with a width of $a\sqrt{3}/2$, where $a$ is the PWC separation. In the vertical case, the width of each OD stripe is $3a/2$ and the PWCs are placed $a/4$ either side of the midline.
Previous studies \citep{bartfeld_relationships_1992,hubener1997spatial,obermayer_geometry_1993} revealed that PWCs tend to lie near the midlines of  OD stripes. More precisely, the mean distance between the midline of the OD stripe and the nearest PWCs is around 100 $\mu$m in cat, which is around $1/5$ of the single OD stripe width \citep{crair1997ocular}. Hence, the latter of the above alternatives better matches with the experiments, a point we consider further below.
Additionally, the OP-OD topology of our hexagonal lattice also closely matches with previous findings \citep{muir_embedding_2011,paik2011orientation_hexa}, which have predicted (also have justified on experimental OP maps) that columns with the same OP are approximately arranged in perturbed hexagonal lattice (i.e., not perfectly regular).

In both square and hexagonal cases, the distances between pinwheels with opposite signs are shorter than the distances between PWCs of the same sign, which matches with the experimental findings in \citep{muller_analysis_2000}. Additionally, the OP-OD topologies in the two cases would not be affected if we stretch or compress the polygons in the vertical or horizontal direction.

\subsection{Possible OP-DP Topologies}
\label{sec:op_dp_topology}
We now discuss constraints on the combined OP-DP map both locally and across square and hexagonal lattices.

As noted above, OP for oriented edges strongly constrains DP, because only the component of motion perpendicular to an edge is detected, at least by simple cells in V1, so DP and OP angles are always perpendicular  \citep{kisvarday2001calculatingDP,shmuel1996functional, swindale1987surface,swindale2003spatialpattern,weliky1996DPsystematic}. An important difference exists because OP only has a range of 180$^{\circ}$ because rotation of an edge through this angle leaves it unchanged; in contrast, DP covers the full 360$^\circ$ range of possible directions of motion of edges. 
\begin{figure}[ht!]
\includegraphics[width=0.4\textwidth]{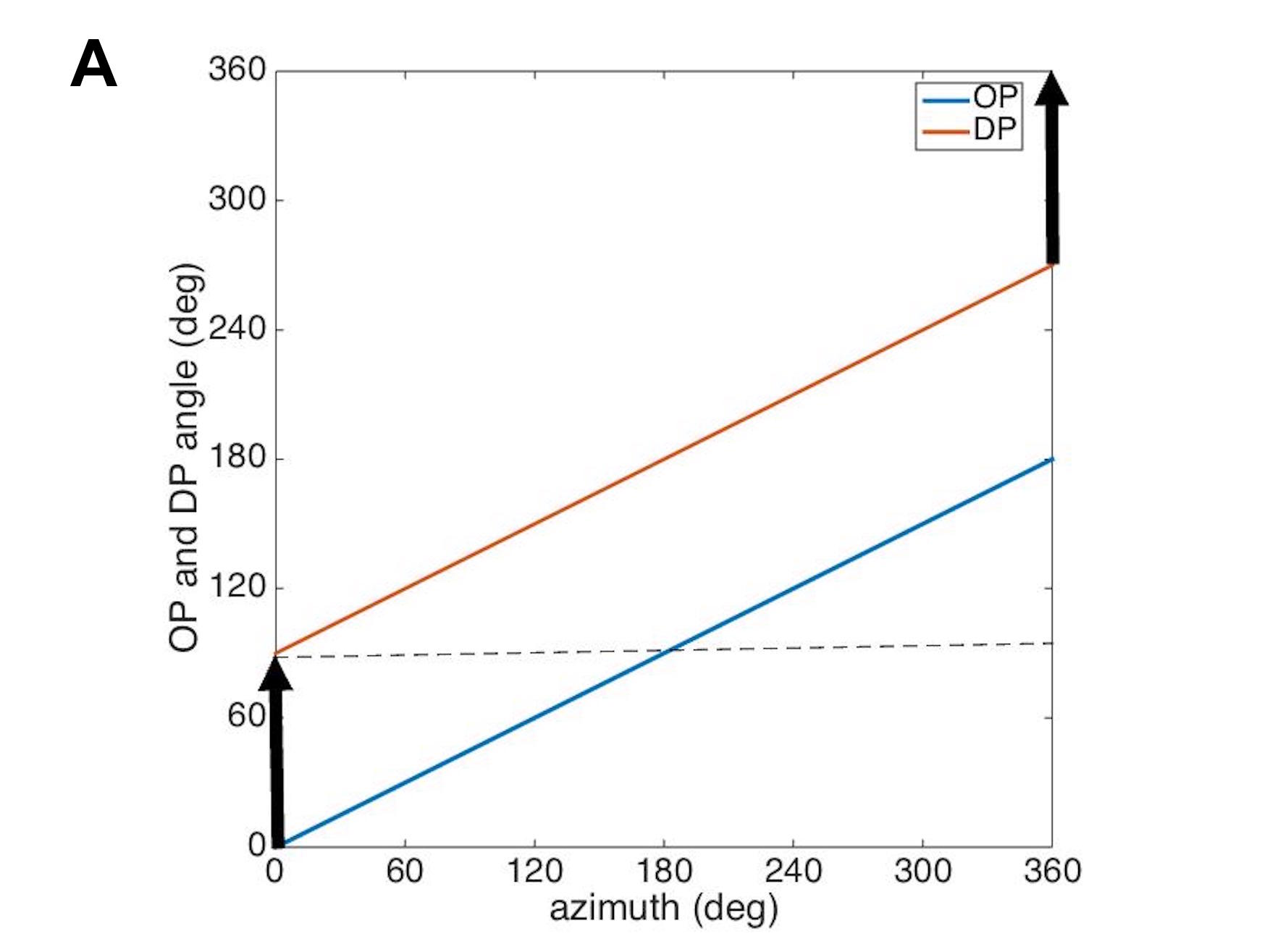}
\includegraphics[width=0.3\textwidth]{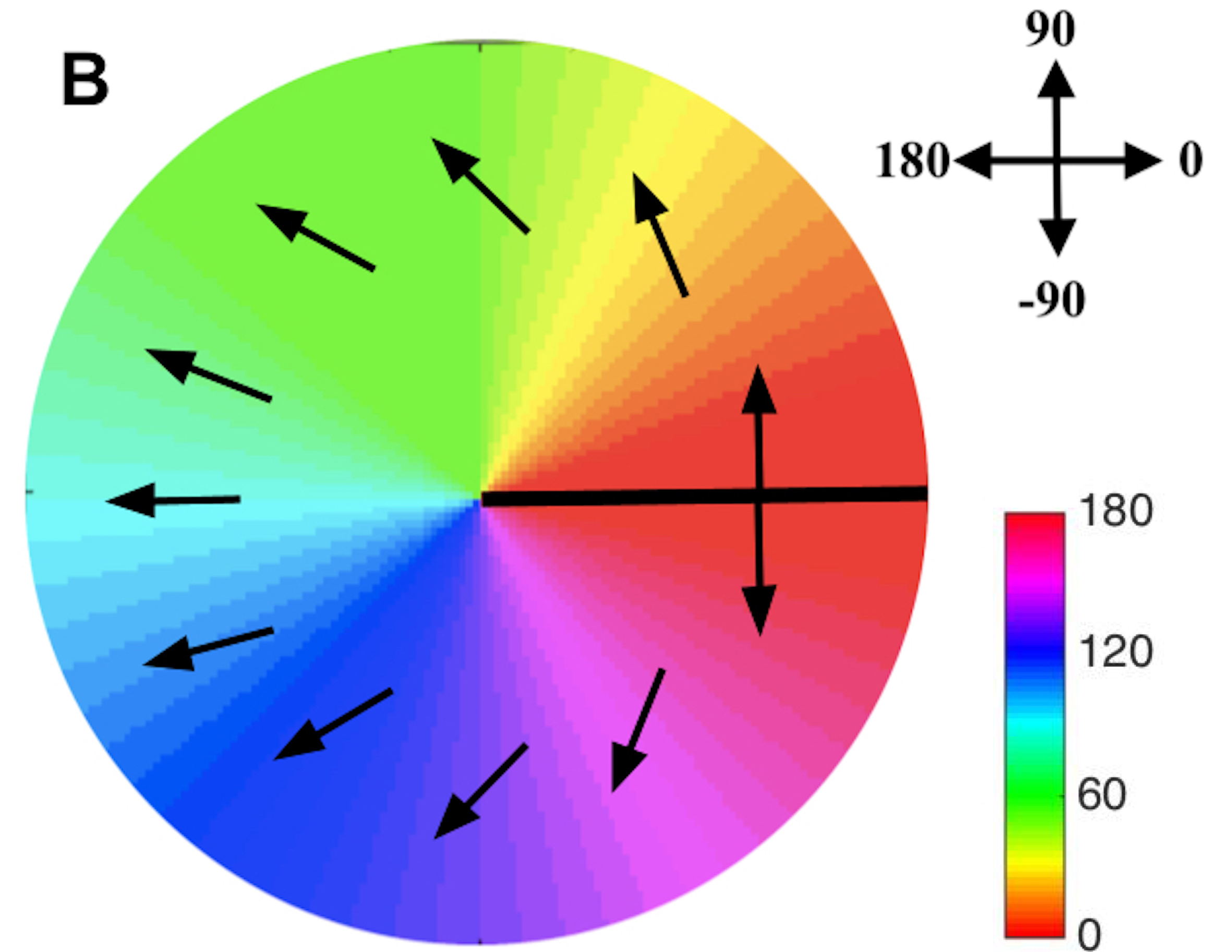}
\includegraphics[width=0.4\textwidth]{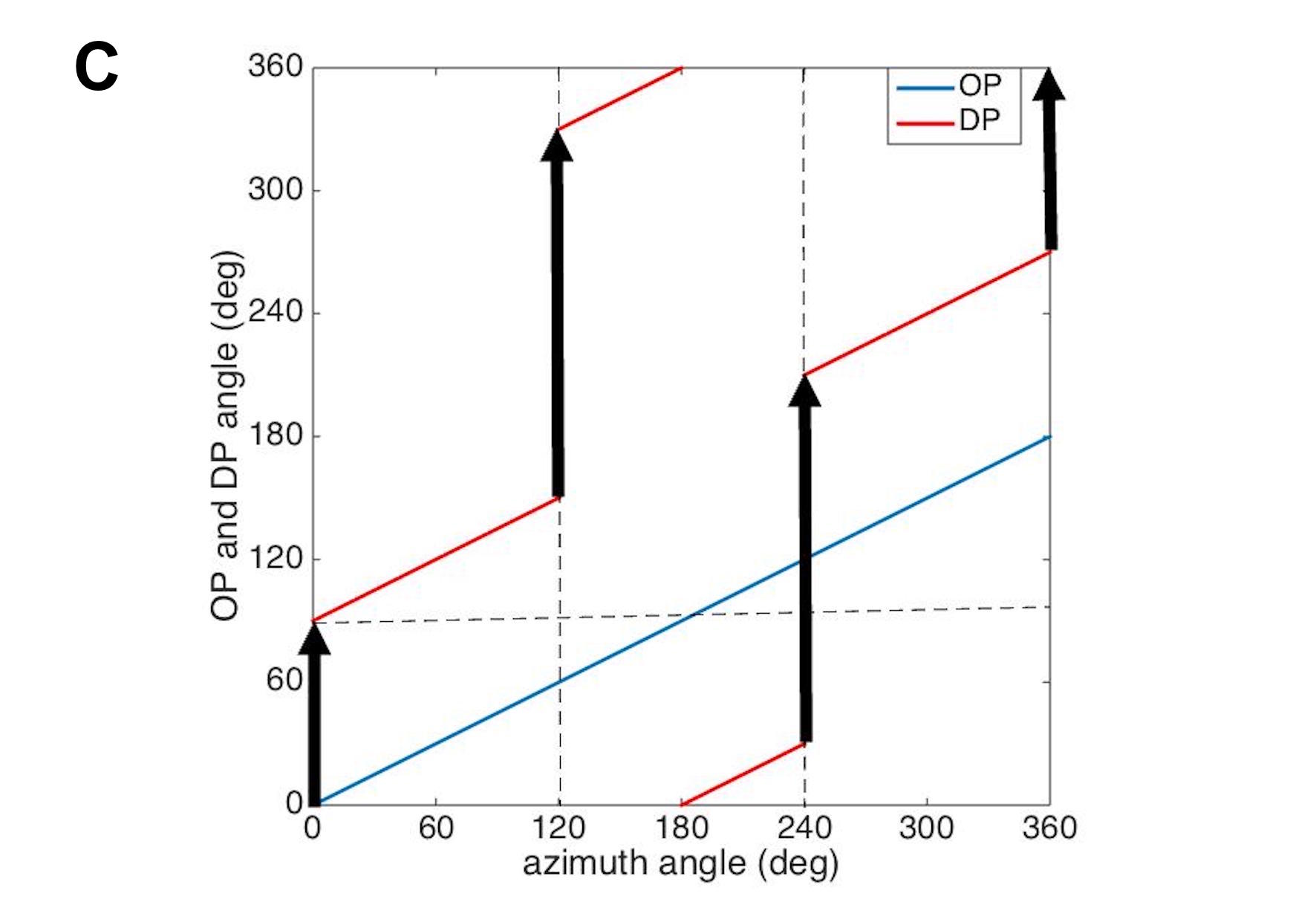}
\includegraphics[width=0.30\textwidth]{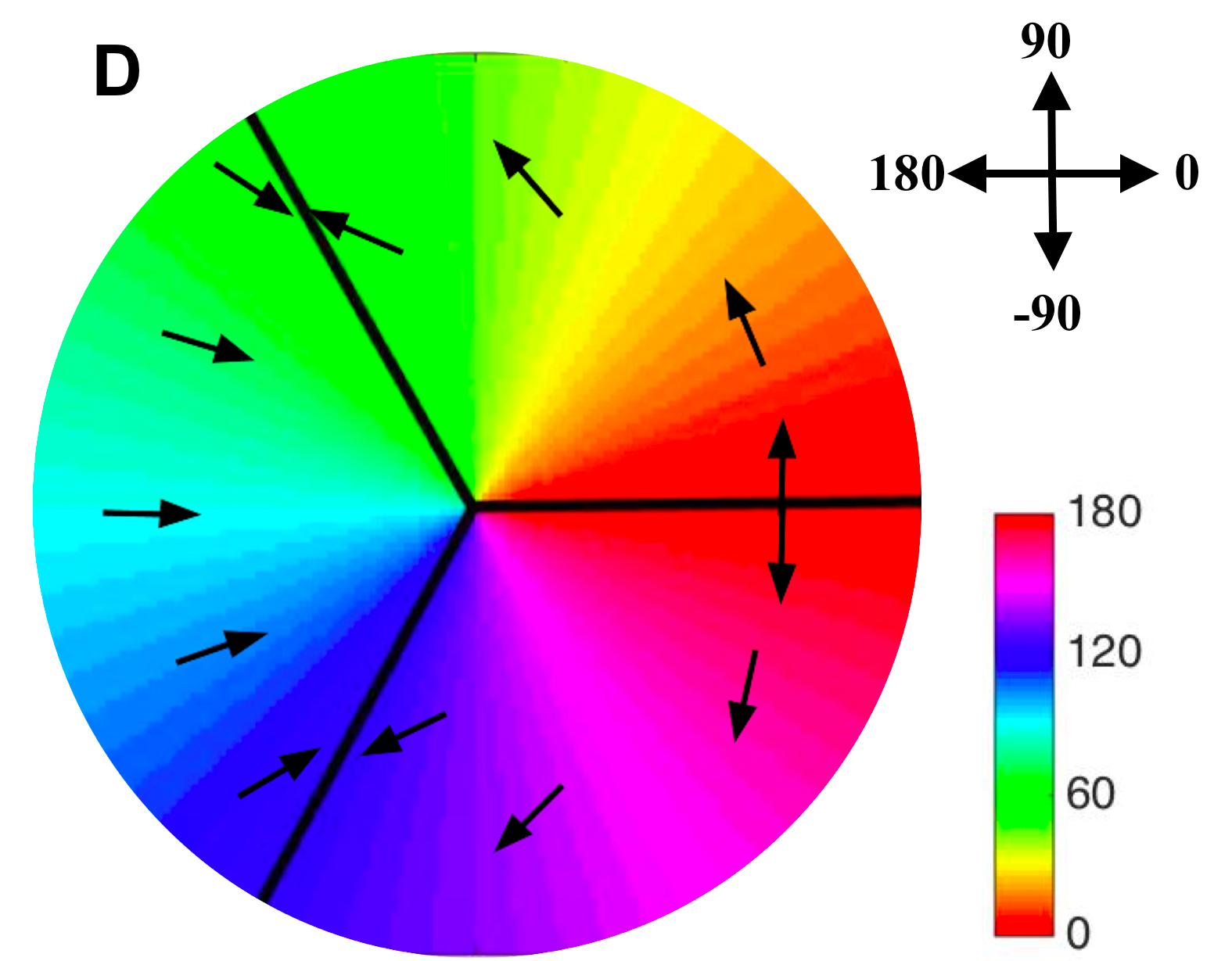}
\caption{\label{fig:op_dp_variation_with_azimuth_angle} Fracture geometry near a positive PWC. \textbf{(A)} OP (blue) and DP (red) vs.~azimuth around the PWC showing one fracture as a vertical arrow. \textbf{(B)} OP-DP map corresponding to frame \textbf{(A)}, with arrows indicating DP and colors showing OP. \textbf{(C)} As for \textbf{(A)} but with three equally spaced fractures. \textbf{(D)} OP-DP map corresponding to frame \textbf{(C)}.}
\end{figure}

\begin{figure}[ht!]
\includegraphics[width=0.35\textwidth]{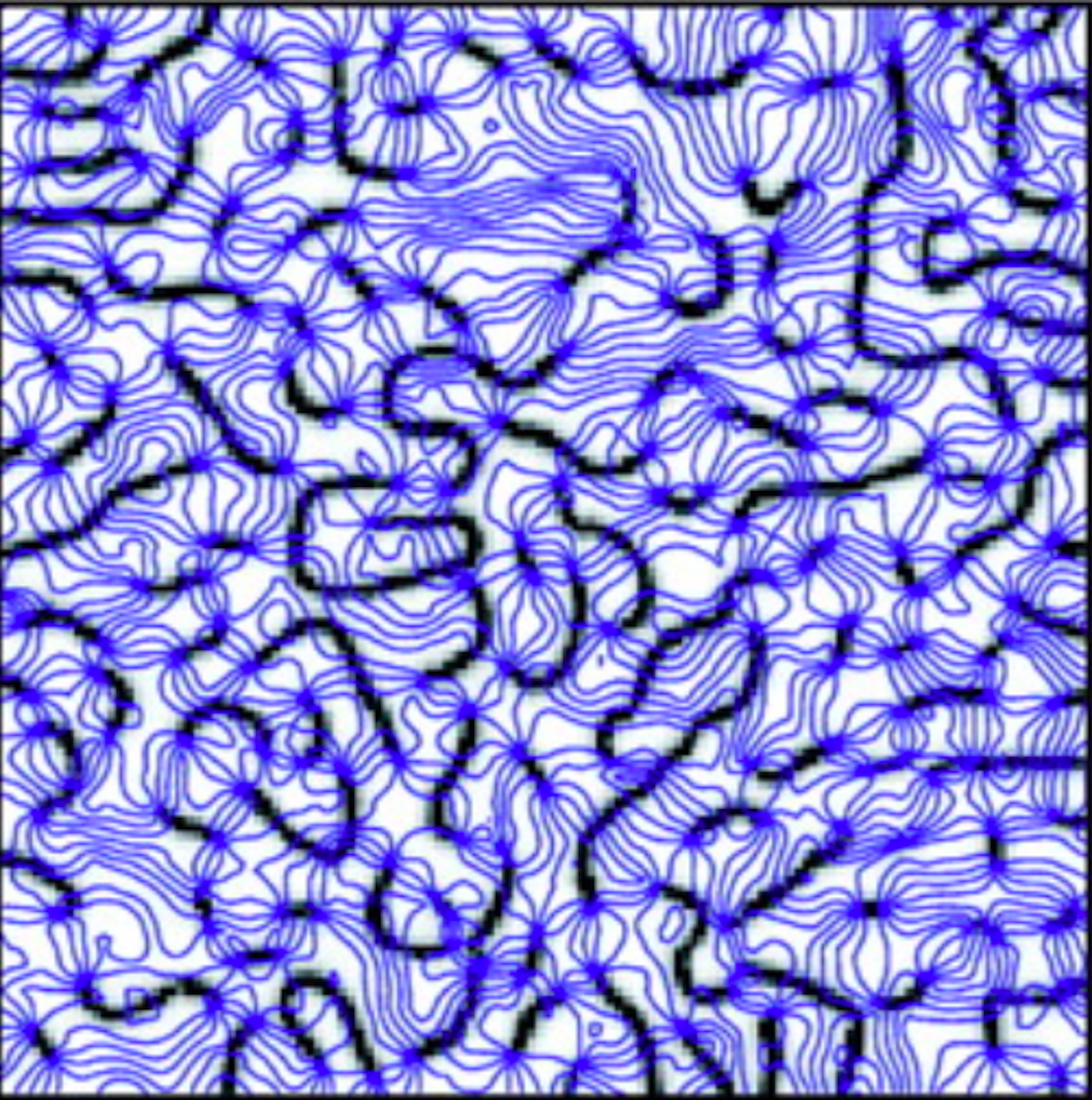}
\caption{\label{fig:simulated_DP_fracture_on_OP} Simulated map showing the relationship between OP map and DP fractures. Reprinted with permission from \cite{carreira_computation_map_2004}. The OP is plotted by blue contour lines and the DP fractures are in black.}
\end{figure}

Figure~\ref{fig:op_dp_variation_with_azimuth_angle}(A) shows the variation of OP vs.~azimuth around a positive PWC, starting at 0$^\circ$ and increasing linearly to $180^\circ$, which is equivalent to 0$^\circ$. (Note that all angles can be globally incremented by any fixed amount without changing our arguments.) If DP correspondingly starts at $90^\circ$ it reaches $270^\circ$ (i.e., covers half of the full DP cycle of $180^\circ$) after a complete circuit around the PWC, meaning that it is pointing in the opposite direction and there must be a discontinuity of 180$^\circ$ at that point --- a DP fracture --- as shown in Fig.~\ref{fig:op_dp_variation_with_azimuth_angle}(B). Although the fracture could equally well occur at some other OP angle, its existence is inescapable. Figures~\ref{fig:op_dp_variation_with_azimuth_angle}(C) and (D) show another possibility, in which three fractures occur during the circuit around the PWC. In general, the total rotation of DP during a circuit of a PWC must be an integer multiple of 360$^\circ$, so if there are $n$ fractures, we must have
\begin{equation}
\label{eq:op_dp_cycle}
    180^\circ(1+n)=360^\circ m,
\end{equation}
where $m$ is an integer. This equation can only be satisfied for odd $n$, which confirms that at least one fracture must emanate from each PWC,  consistent with statistics of experimental DP maps that show that more than $90\%$ of PWCs originate one or more observable DP fractures (most commonly 1 or 3), and most of these end at nearby PWCs \citep{shmuel1996functional,swindale2003spatialpattern}. In addition, simulations of multiple feature maps \citep{carreira_computation_map_2004,Miikkulainen_visual_maps} also predict the existence of DP fractures originating from PWCs, as shown in Fig.~\ref{fig:simulated_DP_fracture_on_OP}. Moreover, this figure clearly shows that almost all DP fractures connect two neighboring PWCs.  

\begin{figure}[ht!]
\includegraphics[width=0.45\textwidth]{ 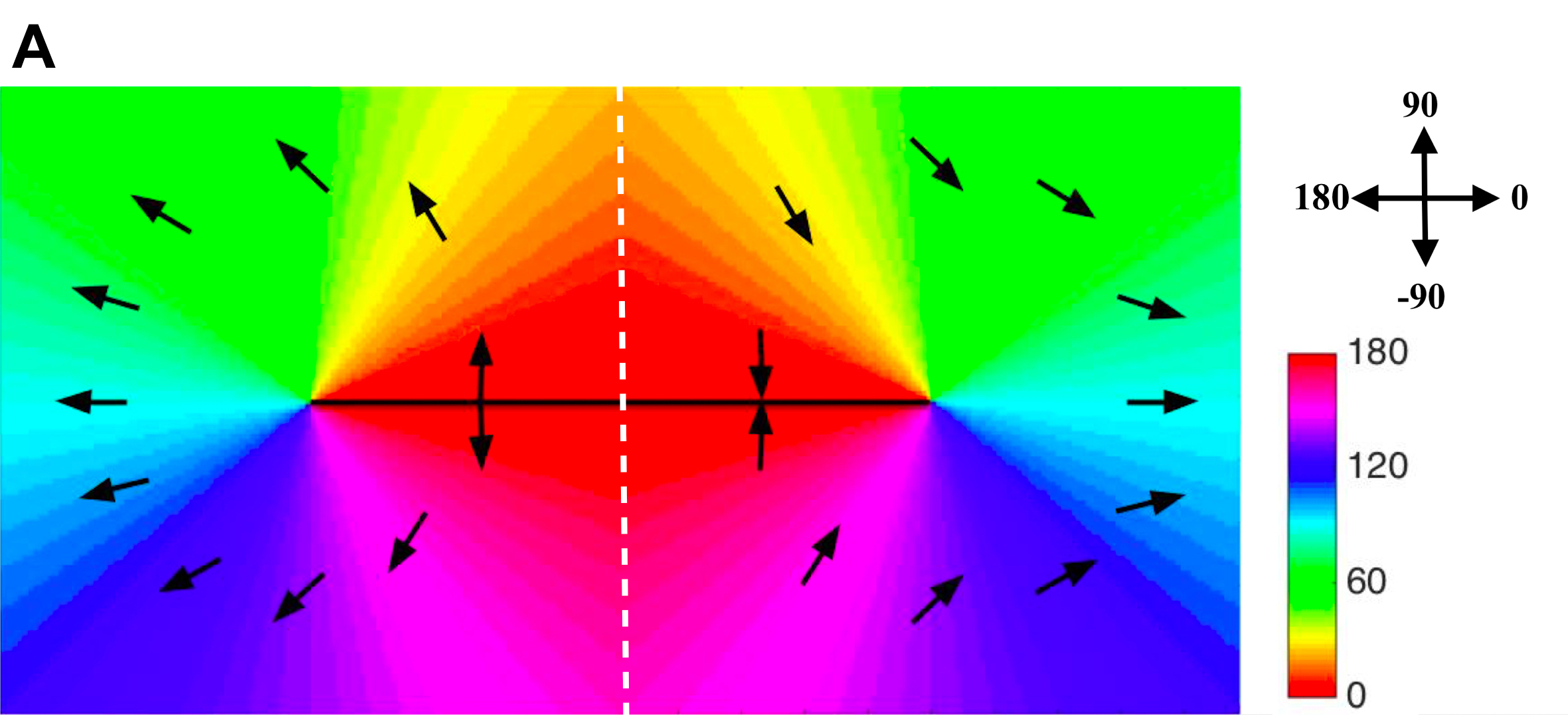}
\includegraphics[width=0.45\textwidth]{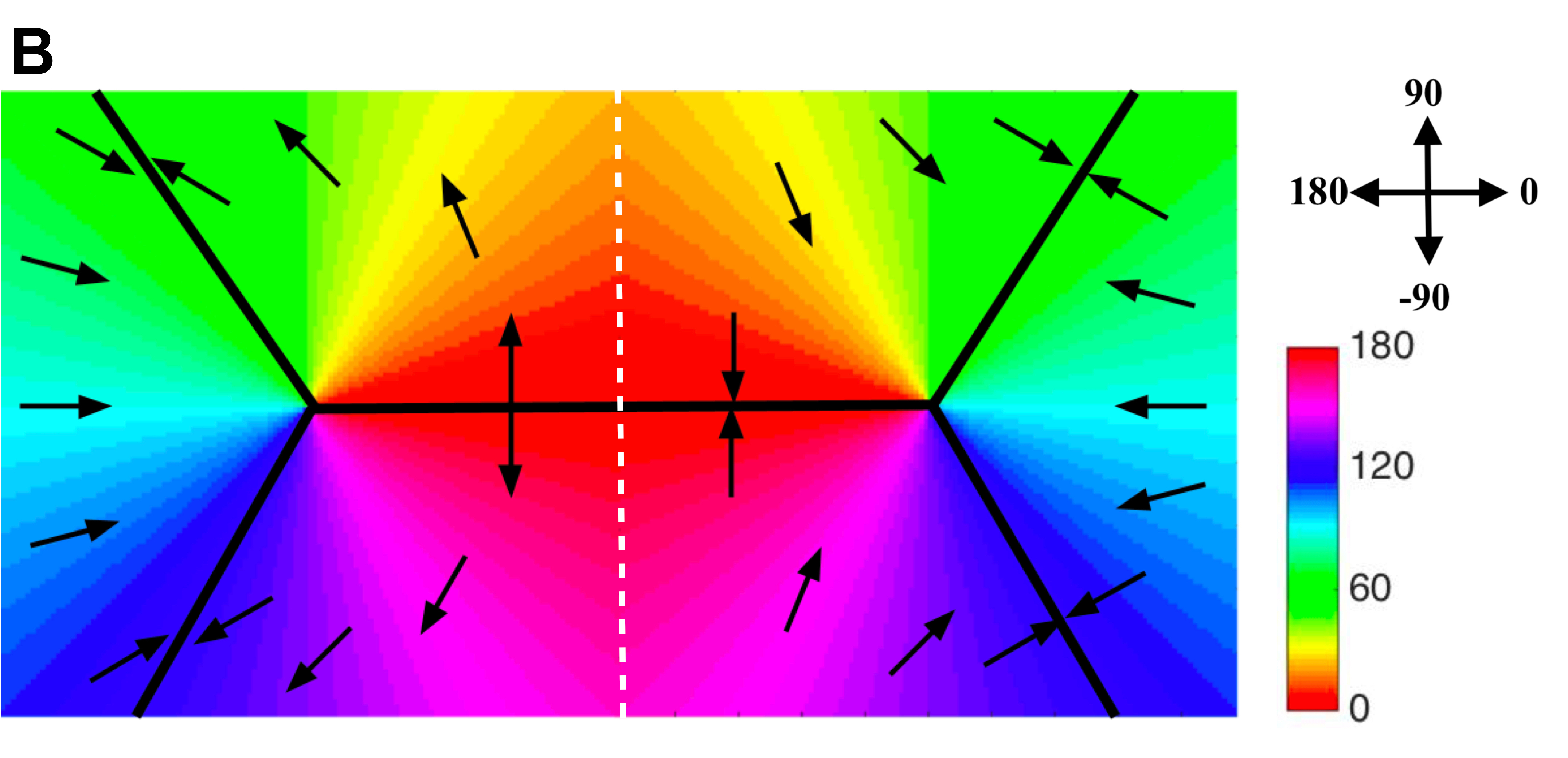}
\caption{\label{fig:DP_coverage_with_two_pw} Schematic of how full DP coverage is achieved between adjacent PWCs linked by a fracture. The solid black line represents the DP fracture, the black arrows show the DP, and the color bar indicates the OP in degrees. The white dashed lines indicate loci of zero DP selectivity. \textbf{(A)} One DP fracture per OP pinwheel. The left pinwheel covers $180^\circ$ of DP ranging from $90^\circ$ to $-90^\circ$, and the right pinwheel covers the rest $180^\circ$ of DP.  \textbf{(B)} Three DP fractures per OP pinwheel. The left pinwheel covers DP ranging from $90^\circ$ to $150^\circ$, $-30^\circ$ to $30^\circ$, and $-150^\circ$ to $-90^\circ$, and the right pinwheel covers the remaining three sectors of DP.}
\end{figure}

We saw in Fig.~\ref{fig:op_dp_variation_with_azimuth_angle} that DP only covers a 180$^\circ$ range around any given PWC, which is also consistent with experiments \cite{swindale2003spatialpattern,weliky1996DPsystematic}. This means that the other PWC in each OD stripe within a hypercolumn must cover the remaining range to preserve symmetry and the definition of the hypercolumn itself. Together with the fact that the neighboring PWC is of opposite sign, this further constrains the local topology of the combined map to be of the form shown in Fig.~\ref{fig:DP_coverage_with_two_pw}(A) for a case of two PWCs linked by a fracture. We see that DP ranges from $90^\circ$ to $270^\circ$ (equivalent to $-90^\circ$) around the positive PWC at left, whereas the remaining range of $-90^\circ$ to $90^\circ$ is covered around the negative PWC at right. Likewise, the case of 3 DP fractures per PWC in Fig.~\ref{fig:DP_coverage_with_two_pw}(B) also covers the full range of DP, with each pinwheel covering half of the range. The left pinwheel covers three intervals of DP, from $90^\circ$ to $150^\circ$, $-30^\circ$ to $30^\circ$, and $-150^\circ$ to $-90^\circ$, while the remaining DP intervals are covered by the right pinwheel. 

The topology seen in Fig.~\ref{fig:DP_coverage_with_two_pw} implies the existence of important additional features of the DP map: at the midpoint (or some other point in a less idealized case) of the fracture, regions of opposite DP abut on the same side of the fracture. This can only occur if the DP selectivity vanishes at the midpoint, akin to the vanishing of OD selectivity at the boundary between L and R stripes and of OP selectivity at PWCs. A one-dimensional locus of vanishing DP selectivity must then radiate in both directions from the midpoint; in our idealized case, this will be a straight line, which is shown as white dotted line in Fig.~\ref{fig:DP_coverage_with_two_pw}, which represents a further DP fracture.

\begin{figure}[ht!]
\includegraphics[width=0.35\textwidth]{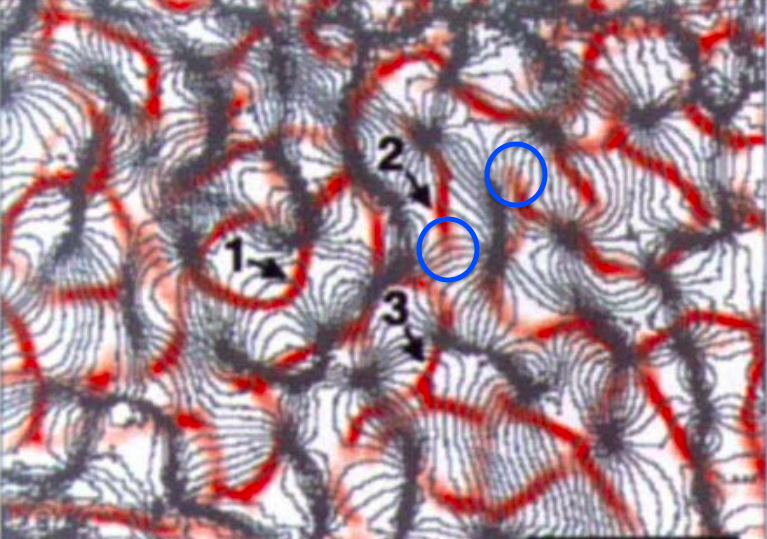}
\caption{\label{fig:DP_fracture_fade_out} Experimental map showing DP fractures (red) on OP contours (black), adapted with permission from \citep{weliky1996DPsystematic}. The numbered sites indicate some places where DP fractures intersect OP iso-orientation contours at steep angle. The blue circles highlight some locations where DP fractures fade out around their midpoints.}
\end{figure}

The new constraints noted in the previous paragraphs are consistent with experimental features seen in Fig.~\ref{fig:DP_fracture_fade_out}: (i) the radiation of DP fractures from PWCs, parallel to OP contours; (ii) the apparent ``fading out'' of many DP fractures between PWCs, examples of which are indicated by the blue circles, with some apparently resuming in much the same direction but opposite DP; and (iii) the sudden $\sim90^\circ$ changes in direction of DP fractures, with many crossing OP contours at steep angles.

The simulated DP maps in Fig.~\ref{fig:simulated_DP_fracture_on_OP} do not show the fading-out of fractures, a difference that may be due to the ability of simulations to determine DP reversals even where DP sensitivity is to low to be observed experimentally. However, simulated DP fractures do tend to follow OP contours, or to intersect them roughly perpendicularly, consistent with our predictions.

We now consider OP-DP maps when PWCs are arranged in a square or hexagonal lattice. In a square lattice the only way in which an odd number of fractures can emanate from each PWC without any ending between PWCs is for there to be one per PWC, but this cannot be achieved while preserving the fourfold symmetry of the lattice and very often three fractures are seen to emerge from a PWC \citep{kisvarday2001calculatingDP,swindale1987surface,swindale2003spatialpattern}, so we do not consider this case further. 

The symmetry of the hexagonal lattice is consistent with there being three fractures emanating from each PWC and connecting it to its nearest neighbors, which are of opposite sign, as seen in Fig.~\ref{fig:hypercol_grid}(C). Topology alone does not force the fracture to parallel the OP contours, but the requirement that neighboring pinwheels cover all possible DPs means that fractures are directed toward one another along the same contour as they emerge from the PWCs, so contour following is favored. In the most symmetric case, DP selectivity would vanish along lines perpendicular to the midpoints of each edge of the hexagons, and would meet at the hexagon centers.  Hence, we predict that the midpoints will have low sensitivity, consistent with the apparent ``fading out'' commonly seen experimentally. We also predicte the centers of hexagons to have low DP selectivity, but have not yet been able to locate any experimental results that would enable this prediction to be tested.

\subsection{Possible OD-OP-DP Topologies}
Having found that the hexagonal lattice is the only possible arrangement for our idealized OD-OP-DP map, we now consider how the OD bands relate to the OP-DP map. Figure~\ref{fig:hypercol_grid}(C) shows the two topologically distinct arrangements. We have already noted in Sec.~\ref{sec:topology}A that the case shown with vertical OD stripes has a better match with the spread of PWCs relative to the midlines of OD stripes than the case shown with horizontal OD stripes. We now further evaluate these alternatives against experimental data on PWC density and layout within OD stripes. 

Let us denote the spacings between PWCs of the same sign as $\lambda_\parallel$ and $\lambda_\perp$ in directions parallel and perpendicular to the OD boundaries, respectively, as shown in Fig.~\ref{fig:hypercol_grid}(C). Results derived from the geometry of the two hexagonal cases are summarized in Table \ref{table:distance_to_OD}, together with corresponding experimental findings from macaque monkey \cite{obermayer_geometry_1993}. Experimentally, $\lambda_\perp>\lambda_\parallel$, whereas both our idealized arrangements exhibit the reverse inequality. The experimental results thus correspond to the hypercolumns being compressed in the parallel direction in macaque relative to the idealized case, while leaving the topology unchanged. However, the vertical OD arrangement in Fig.~\ref{fig:hypercol_grid}(C) gives by far the better agreement with the observed $\lambda_\perp/\lambda_\parallel$, being only $26\pm7$\% smaller. Because it also agrees better with the spread of PWCs relative to the OD stripe midlines \citep{crair1997ocular}, as discussed above, we argue that it is the best option among the symmetric maps considered. For convenience below we write the ratio of the experimental value of $\lambda_\parallel/\lambda_\perp$ to our theoretical prediction of this quantity
\begin{equation}
\label{eq:ratio_lambda}
    \Delta=\frac{0.85}{2/\sqrt{3}}\approx 0.74.
\end{equation}

\begin{table}[ht!]
\caption{\label{table:distance_to_OD} Spacings of nearest PWCs of the same sign, $\lambda_\parallel$ and $\lambda_\perp$, parallel and perpendicular to the OD boundary, respectively, for the horizontal and vertical OD stripe  orientations in Fig.~\ref{fig:hypercol_grid}(C), and for experimental data of macaque monkey from \citep{obermayer_geometry_1993}.}
\begin{tabular}{|p{0.11\linewidth} |p{0.3\linewidth} |p{0.27\linewidth} |p{0.23\linewidth}|}
\hline
             &\textbf{Horizontal OD arrangement}&\textbf{Vertical OD arrangement}& \textbf{Experiment} (mm) \\ \hline
$\lambda_{\parallel}$ & $3a$            & $a\sqrt{3}$     & $0.64\pm 0.02$  \\ \hline
$\lambda_{\perp}$ & $a\sqrt{3}/2$         & $3a/2$       & $0.76\pm 0.04$ \\ \hline
$\lambda_{\parallel}/\lambda_{\perp}$ & $\sqrt{3}\approx 1.73$ & $(2/\sqrt{3})\approx 1.15$ & $0.85\pm0.08$ \\ \hline
\end{tabular}
\end{table}

The width of a single OD stripe equals $\lambda_\perp$. If we choose $a=2\lambda_\perp/3$, so that $\lambda_\perp$ matches the experimental width, and set $\lambda_\parallel/\lambda_\perp$ to 0.85 to accord with experiment, this gives the values of $a$ shown in the forth column in Table~\ref{tab:average_pwc_density}.

We next use our hexagonal array to predict the average pinwheel density and compare it with results from experiments. In our vertical OD arrangement, the pinwheel density is 
\begin{eqnarray}
\label{eq:pwc_density}
    \rho&=&\frac{4}{3\sqrt{3}a^2\Delta}\approx \frac{0.77}{a^2\Delta}.
\end{eqnarray}
where $a$ is the side length of the hexagon, and $\Delta$ is the ratio from Eq.~(\ref{eq:ratio_lambda}), which is included here to accommodate the effect of real OP map being compressed in the direction parallel to the OD stripes. The value of $a$ in the sixth column of Table~\ref{tab:average_pwc_density} is calculated from Eq.~(\ref{eq:pwc_density}) with the experimental pinwheel density shown in the second column. 

\begin{table*}[!ht]
\caption{\label{tab:average_pwc_density} V1 pinwheel density (pinwheels mm$^{-2}$), average OD stripe width ($\lambda_\perp$), and the mean periodicity of OP ($\Lambda$) from various experiments. And the pinwheel spacing $a$ of the hexagonal lattice that is derived from these three data sets.}
\begin{tabular}{|l|l|l|l|l|l|l|}
\hline
                 &$\rho$ (mm$^{-2}$)&$\lambda_\perp$ (mm)&$\Lambda$ (mm)&$a$ from $\lambda_\perp$ (mm)&$a$ from $\rho$ (mm)&$a$ from $\Lambda$ (mm)\\ \hline
Human \cite{yacoub_human_op_map}           &2.24 &$1.17\pm0.08$&$1.43\pm0.12$&$0.78\pm0.05$&$0.68$&$0.65\pm0.05$\\ \hline
Macaque \cite{ho2021orientation_murinus,obermayer_geometry_1993,obermayer1997singularities}& $8.1$ &$0.41\pm0.03$&$0.695$&$0.27\pm0.02$&$0.36$&$0.32$\\ \hline
Galago \cite{ho2021orientation_murinus,Kaschube_PWC_density,xu2005functional_map_galago} &$6.64$ & $0.53\pm0.12$ &$0.68$& $0.35\pm0.08$ &$0.39$& $0.31$\\ \hline
Ferret \cite{Kaschube_PWC_density,keil2012response_to_universality,law1988organization_ferret} & $4.16\pm0.11$ & $0.41\pm0.14$ &$0.87$&$0.28\pm0.09$ & $0.50\pm0.01$&$0.40$ \\ \hline
Cat \cite{Anderson_cat_Od,HUBENER2002131,keil2012response_to_universality,levay_cat_Od} &$2.5\pm0.5$&$0.6\pm0.1$ &$1.01$ &$0.40\pm0.07$& $0.66\pm0.06$& $0.46$ \\ \hline
Tree shrew \cite{bosking_orientation_1997,ho2021orientation_murinus}    & $9.6$ &---&$0.62$ & ---                   &  $0.33$& $0.28$\\ \hline
Mouse lemur \cite{ho2021orientation_murinus} &$10.8$ &--- &$0.54$& ---  &  $0.31$ &$0.25$ \\ \hline
\end{tabular}
\end{table*}

Several points are worth noting in Table~\ref{tab:average_pwc_density}: (i) the value for humans is the experimental one in a region at $5^\circ$ eccentricity (i.e., the angular distance from the center of the visual field) \citep{yacoub_human_op_map}, and the pinwheel densities for other species are averaged values over the whole measurement area in each reference; (ii) cats and ferrets have patchy OD bands, rather than stripes; the OD width for these two animals is rather the average diameter of the OD patches; (iii) ocular dominance stripes are not found in tree shrew and mouse lemur, so the corresponding OD width does not exist.

By comparing the fifth and sixth columns of  Table~\ref{tab:average_pwc_density} we conclude that the estimates of $a$ from human, macaque and galago OD widths and PWC densities are similar in each case, although not always to within the indicated uncertainties. This implies that our hexagonal lattice is a reasonable semiquantitative  approximation to the spatial structure of the experimental OP and OD maps in these three species. For ferret and cat the $a$ values estimated from OD stripe width are smaller than those estimated from pinwheel density. The possible reason for the mismatch is that the OD pattern in these animals appears as elongated patches with variable width and length, rather than true stripes. Although some patches are connected and form short parallel bands in a few millimeters, the overall OD pattern is more irregular than the striped OD arrangement in macaque \citep{Anderson_cat_Od,crair1997ocular,HUBENER2002131,law1988organization_ferret}, so the poorer agreement is not surprising.

\begin{figure}[ht!]
\includegraphics[width=0.37\textwidth]{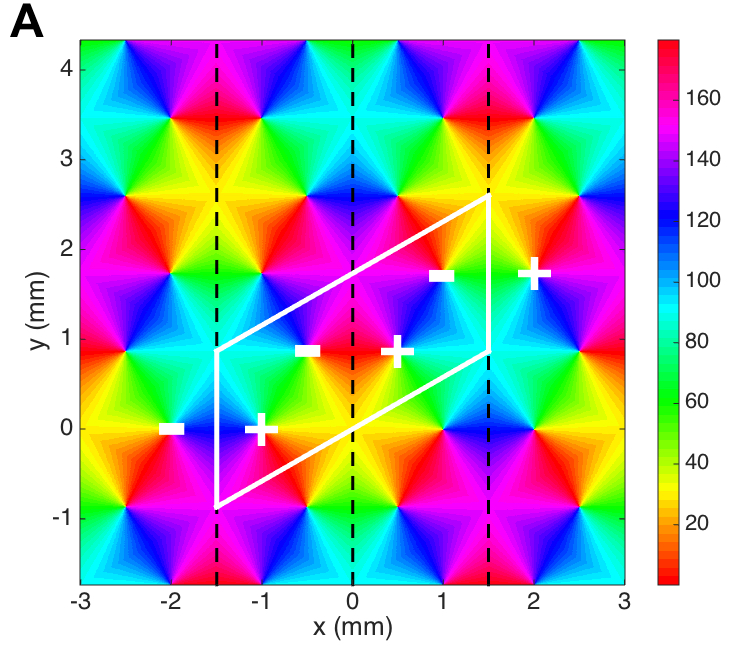}
\includegraphics[width=0.37\textwidth]{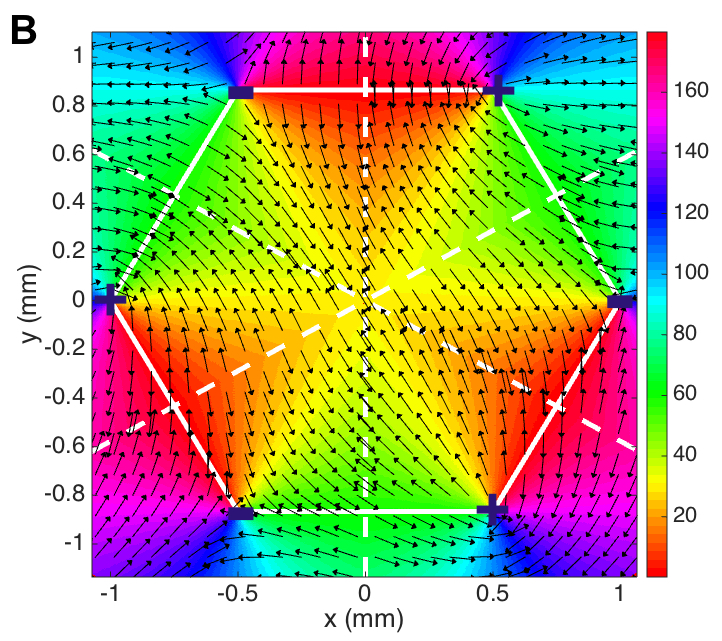}
\includegraphics[width=0.37\textwidth]{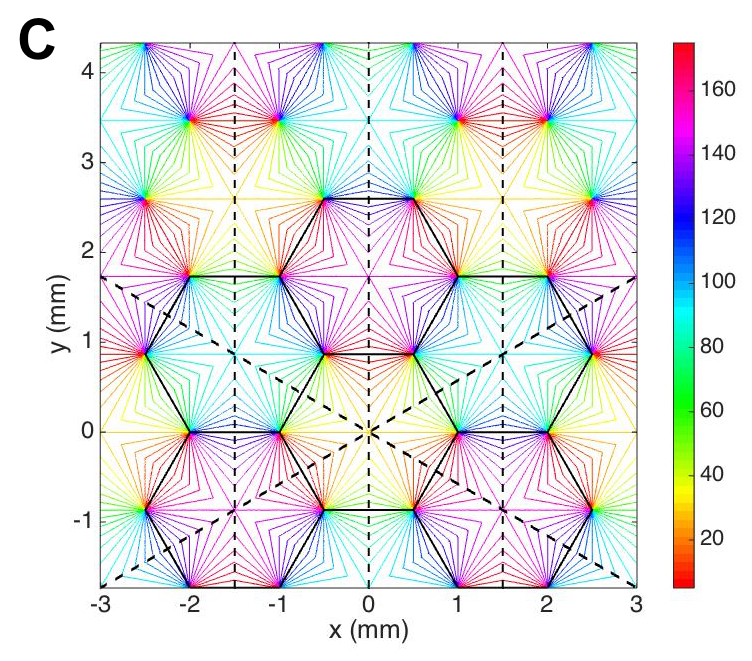}
\caption{\label{fig:op_od_hexa_lattice} Combined feature maps. \textbf{(A)} OP-OD map corresponding to the hexagonal lattice with vertical OD stripes as shown in Fig.~\ref{fig:hypercol_grid}(C). Positive and negative PWCs are marked by $+$ and $-$ signs. Black dotted lines denote the OD borders and the white lines outline one hypercolumn. The color bar indicated the orientation preference in degrees. \textbf{(B)} OP-DP map of one hexagon extracted from (A). OP is color coded and DP is shown by arrows. The white lines indicate DP fractures and the dashed lines indicate zero DP selectivity. Positive and negative PWCs are marked. (C) Examples of DP fractures that parallel OP contours (solid lines) and intersect OP contours (dashed lines) overlaid on colored OP contours. The color bar indicates the OP in degrees.}
\end{figure}

Figure~\ref{fig:op_od_hexa_lattice}(A) shows a color coded OD-OP map corresponding to the hexagonal lattice with vertical OD arrangement shown in Fig.~\ref{fig:hypercol_grid}(C). It is obtained by finding the OP angle  $\phi_{1}$ around one PWC from \citep{liu2021analytic}
\begin{equation}
\phi_{1}=\tan\left(\frac{y-y_c}{x-x_c}\right),
\end{equation} 
where $(x_c,y_c)$ are the coordinates of the PWC, then reflecting the resulting pattern across axes of symmetry to obtain the OP near neighboring pinwheels. The PWCs are located at the vertices of each hexagon, six of which are marked $+$ (for positive pinwheels) and $-$ (for negative pinwheels). The white border outlines a hypercolumn with four pinwheels and the dashed lines are the OD borders. Figure~\ref{fig:op_od_hexa_lattice}(B) shows DP arrows overlaid on an OP map of one hexagon extracted from frame (A). DP reverses direction across fractures (white lines), and loci of zero DP selectivity are marked by dashed white lines. Figure~\ref{fig:op_od_hexa_lattice}(C) shows DP discontinuities overlaid on OP contours to parallel the experimental results in Fig.~\ref{fig:DP_fracture_fade_out} (i.e., the numbered sites). This emphasizes the tendency for discontinuities to either follow OP contours or to intersect them at 90$^\circ$.

Equation \eqref{eq:pwc_density} gives the PWC density in terms of the PWC separation in the hexagonal lattice. Some previous works argued for a universal PWC density in {\it aperiodic} arrangements of PWCs, expressed in terms of the mean periodicity $\Lambda$ of OP, independent of species \citep{ho2021orientation_murinus,Kaschube_PWC_density,keil2012response_to_universality}, with 
\begin{equation}
\label{eq:density_experi}
\rho=\frac{\pi}{\Lambda^2}.  
\end{equation}

\begin{figure}[ht!]
\includegraphics[width=0.4\textwidth]{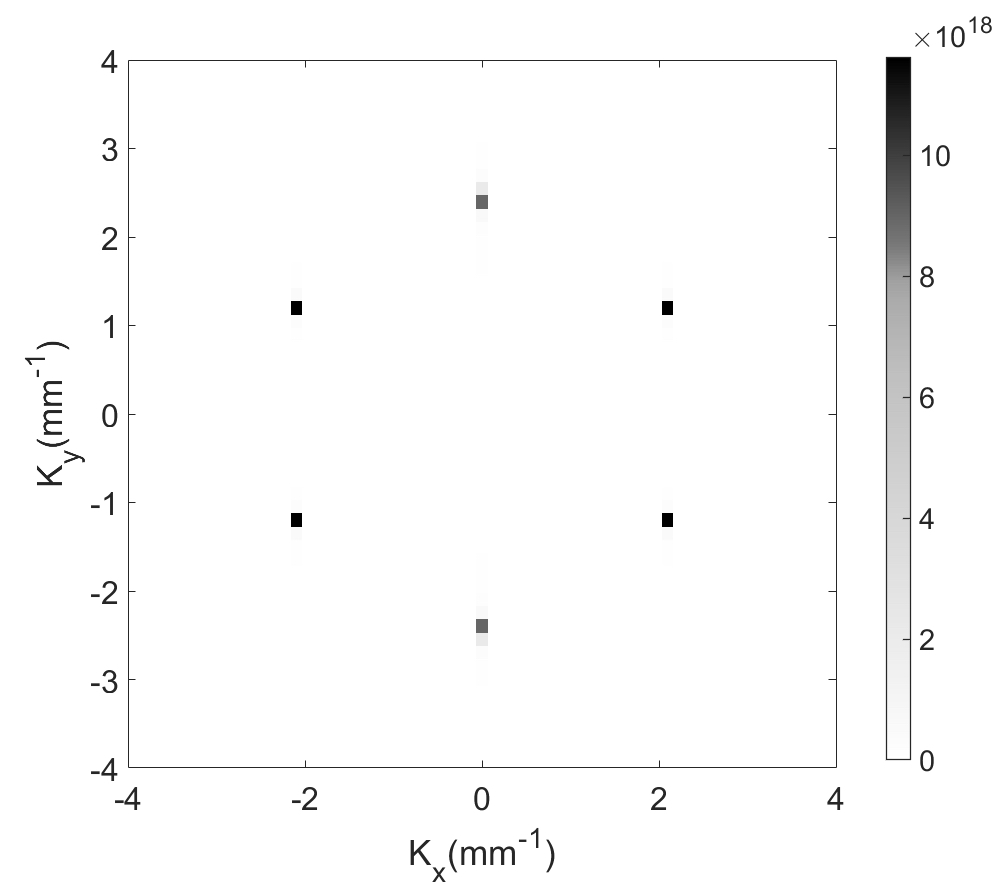}
\caption{\label{fig:mag_plot_hexa} Power spectrum of the hexagonal lattice of OP shown in Fig.~\ref{fig:op_od_hexa_lattice}(A). The color bar indicates the squared amplitude.}
\end{figure}

Equations \eqref{eq:pwc_density} and \eqref{eq:density_experi} are equivalent if 
\begin{equation}
    \Lambda=\left(\frac{3\pi\Delta\sqrt{3}}{4}\right)^{1/2}a\approx 1.86 a;
\label{eq:Lambda_from_eqn_3}
\end{equation}
however, $\Lambda$ is the mean periodicity of OP, so we should estimate it for our hexagonal lattice. We estimate $\Lambda$ via Fourier analysis of the hexagonal lattice in Fig.~\ref{fig:op_od_hexa_lattice}(A) \citep{Kaschube_PWC_density,obermayer_statistical-mechanical_1992,obermayer_geometry_1993}. Firstly, we compute the power spectrum of the hexagonal lattice as the squared amplitude of its Fourier transform, and the resulting spectrum is shown in Fig.~\ref{fig:mag_plot_hexa}. It has six peaks arranged in hexagon and the magnitude of the corresponding $\mathbf{k}$ is
\begin{equation}
\label{eq:original_k}
  k_0=  \frac{4\pi}{3a\sqrt{3}}.
\end{equation}

As we mentioned earlier, the experimental map corresponds to compressing the hexagonal lattice by a factor of $\Delta$ [i.e., Eq.~(\ref{eq:ratio_lambda})] in the vertical direction. In order to take this factor into account, the components of $\mathbf{k}$ in the Fourier domain are stretched by a factor of $1/\Delta$ in the vertical direction. If we denote the stretched $|\mathbf{k}|$ by $k$, the top and bottom $\mathbf{k}$ modes in Fig.~\ref{fig:mag_plot_hexa} have squared wave number 
\begin{equation}
    {k}_{1}^2=\left ( \frac{k_{0}}{\Delta}\right )^2.
\end{equation}
and the other four modes have squared wave number
\begin{equation}
    k_{2}^2=\frac{k_{0}^2}{4}\left ( 3 + \frac{1}{\Delta^2}\right ).
\end{equation}
The averaged value of $k^2$ is then
\begin{equation}
    \left <k^2\right> =\frac{1}{6}\left (2k_{1}^2+4k_{2}^2\right)\\
    =\frac{k_{0}^2}{2}\left (\frac{1+\Delta^2}{\Delta^2}\right).
\end{equation}
The mean periodicity of OP is evaluated via
\begin{eqnarray}
\Lambda &=&2\pi/\sqrt{\left <k^2\right>}, \\
&=&3\sqrt{\frac{3}{2}}\frac{\Delta}{\sqrt{1+\Delta^2}}a,\\
&\approx&2.19a.
\label{eq:Lambda_hexa_approx}
\end{eqnarray}
The estimate of $\Lambda$ from Eq.~(\ref{eq:Lambda_from_eqn_3}) is only 15\% smaller than this estimate, which is acceptable, given that we are dealing with a specific regular lattice, whereas the expression in Eq.~(\ref{eq:density_experi}) was postulated for irregular ones. Moreover, the experimentally measured pinwheel densities had deviations of up to $\sim10\%$ from $\pi/\Lambda^2$, which further supports consistency with Eqs~(\ref{eq:Lambda_from_eqn_3}) and (\ref{eq:Lambda_hexa_approx}). We use the above equation to compute $a$ from the experimental $\Lambda$ listed in the fourth column of Table~\ref{tab:average_pwc_density}. The resulting $a$ values in the last column are similar to the ones obtained from pinwheel density, implying that the hexagonal lattice with vertical OD stripes is also semiquantitatively consistent with these data.

\subsection{Selectivities}
\begin{figure}[ht!]
\includegraphics[width=0.35\textwidth]{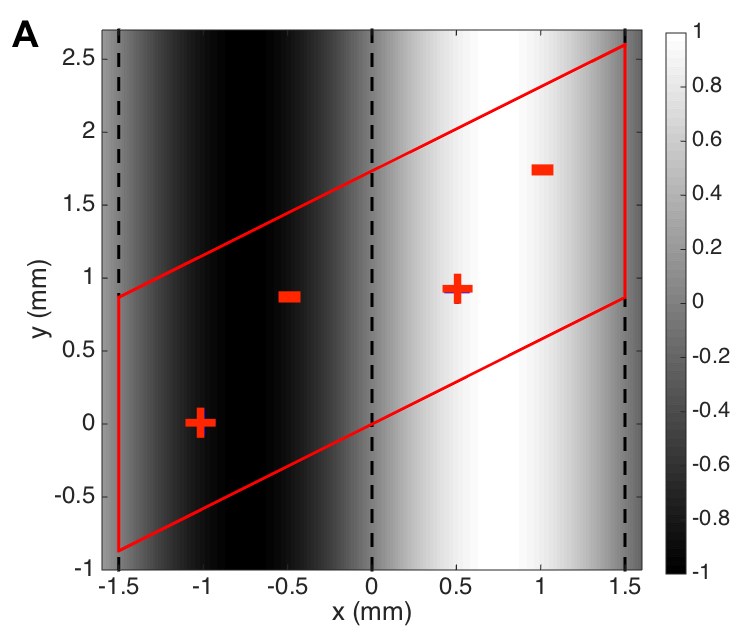}
\includegraphics[width=0.35\textwidth]{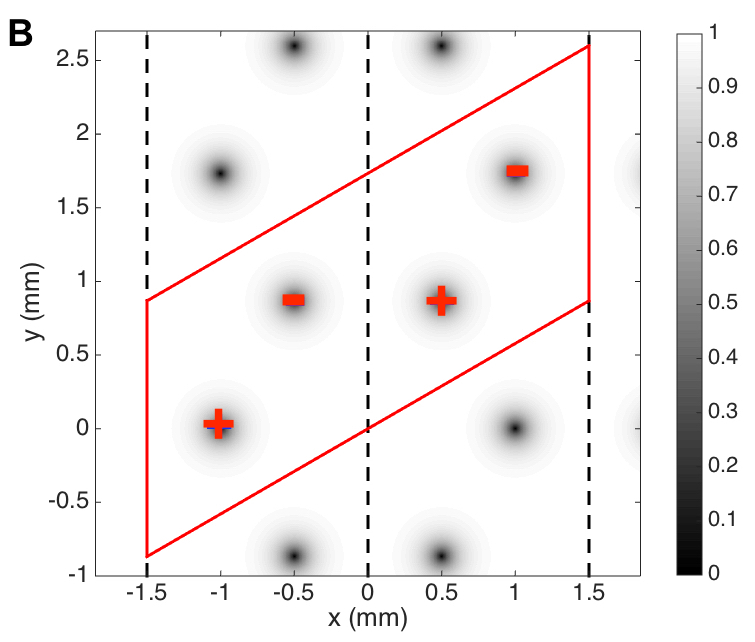}
\includegraphics[width=0.35\textwidth]{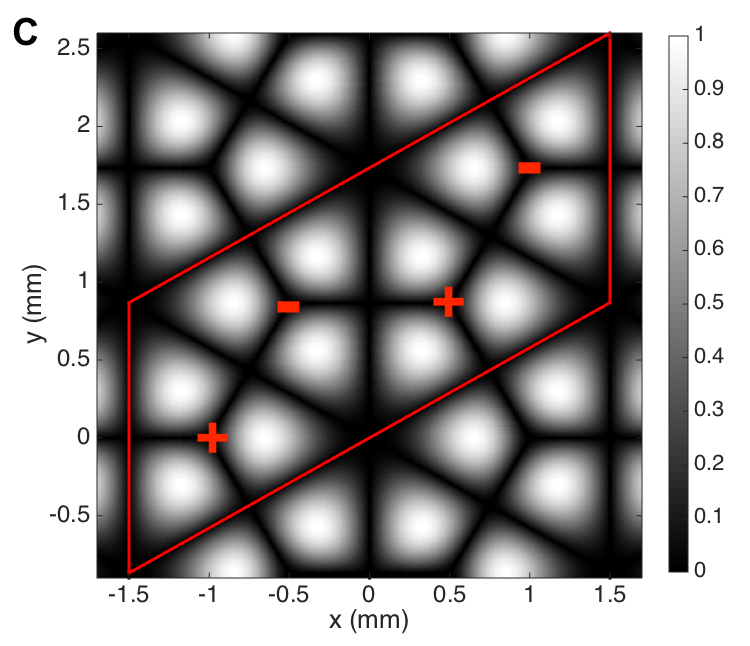}
\caption{\label{fig:schem_selectivity} OD, OP, and DP selectivity in one hypercolumn with PWCs located at the vertices and $a=1$ mm. The red lines outline one hyper column and the dashed lines are the OD border. Four positive and negative OP pinwheels are marked by the $+$ and $-$ signs. \textbf{(A)} OD selectivity map with black and white indicate left and right OD respectively. \textbf{(B)} OP selectivity map with zero selectivity around PWCs and high selectivity elsewhere.  \textbf{(C)} DP selectivity map. Zero selectivity occurs along the DP fractures (i.e., hexagon sides) and the perpendicular bisectors of these fractures. }
\end{figure}

As noted above, the OD, OP, and DP selectivities vary with cortical location and relate to the OD-OP-DP map topology. The main features are:

(i) OD selectivity peaks at the midlines of OD stripes and falls off towards the borders \citep{crair1997ocular,shatz1978ocular}. If we define a net OD selectivity (right minus left) $\nu_{OD}$, normalized to unit maximum, we can approximate $\nu_{OD}$ for illustrative purposes as being sinusoidal with extremums located at the centers of the OD bands:
\begin{equation}
    \nu_{OD} =\sin\left(\frac{2\pi x}{3a}\right),
\end{equation}
where $a$ is the side length of the hexagon. Figure~\ref{fig:schem_selectivity}(A) plots the OD selectivity in one hexagon. Black and white represent left and right OD selectivity, respectively, and the dashed line is the OD border, where $\nu_{OD}=0$ and the overall response is binocular. The centers of OD stripes are located at $x=\pm 3a/4$.

(ii) OP selectivity is low around PWCs, and initially increases roughly linearly with radial distance before saturating after 100-200 $\mu$m   \citep{blasdel_orientation_1992,obermayer_geometry_1993,swindale_review_1996}. We thus approximate the OP selectivity $\nu_{OP}$ for illustrative purposes as
\begin{equation}
    \nu_{OP} =\prod_{i=1}^6\left[1-\exp\left (\frac{-|\mathbf{r}-\mathbf{r}_i|}{r_{0}}\right)\right],
\end{equation}
where the $\mathbf{r}_i$ are the vertices of the hexagon, $r_{0}=80\ \mu$m parametrizes the low OP selectivity around each PWC \citep{maldonado1997orientation,nauhaus2008neuronal,ohki2006highlyorderpinwheel,swindale2003spatialpattern}. The resulting map in Fig.~\ref{fig:schem_selectivity}(B) shows that OP selectivity shows is relatively uniform away from PWCs. It is clear from the figure that the low selectivity regions are around the six pinwheel centers. Furthermore, the selectivity increases as one moving away from the center and reaches maximum in the center area of the hexagon, where OP varies smoothly within this region.

(iii) DP selectivity is low at fractures and reverses sharply across these fractures  \citep{kisvarday2001calculatingDP,shmuel1996functional,swindale2003spatialpattern}. Furthermore, DP selectivity is zero along perpendiculars to the midpoints of DP fractures, as was discussed in Sec.~\ref{sec:op_dp_topology}. We thus approximate the DP selectivity $\nu_{DP}$ as
\begin{equation}
    \nu_{DP} =\prod_{i,j=1}^3 \frac{\lvert\mathbf{r}\cdot {\bf u}_i\rvert}{a} \left|\left(\frac{\mathbf{r}}{a}-\mathbf{u}_i\right)\cdot\mathbf{m}_j\right|\left| \left(\frac{\mathbf{r}}{a}+\mathbf{u}_i\right)\cdot\mathbf{m}_j\right |,
\end{equation}
for illustrative purposes where ${\bf u}_i$ and ${\bf m}_j$ are unit vectors that point from the center of the hexagon toward the three vertices located at $(1/2,\sqrt{3}/2)$, $(1,0)$, and $(1/2,-\sqrt{3}/2)$ and the three edge midpoints at $(0,\sqrt{3}/2)$, $(3/4, \sqrt{3}/4)$, and $(3/4, -\sqrt{3}/4)$. We take the absolute value of the dot product to ensure positive selectivity. This gives the DP selectivity map shown in Fig.~\ref{fig:schem_selectivity}(C), which we have normalized to unit maximum. A few features worth noting are: (i) the DP fractures at hexagon sides have zero selectivity where the DP reverses, consistent with the low selectivity observed experimentally at these fractures  \citep{swindale2003spatialpattern,weliky1996DPsystematic}, given that neurons have nonzero size and measurements have limited spatial resolution; and (ii) the selectivity is also zero at the midlines that are perpendicular to the hexagon sides, so we predict that it should be low in the center of the hexagon where OP sensitivity is high, but we are not aware of any published experiments on this point.

\section{\label{conclusions}Summary and Discussion}
In this study, we have analyzed the possible topologies of the combined OD-OP-DP map by   requiring mutual consistency, periodicity, and maximum symmetry. The main findings are:

(i) We have combined the characteristics of OD, OP, and DP maps from various experiments into a hypercolumn, within which all possible values of these feature preferences occur. In general, each hypercolumn contains parallel left and right OD stripes with a pair of OP pinwheels of opposite sign in each stripe. DP at each location is perpendicular to its OP, and each pinwheel center has an odd number DP fractures originating from it. 

(ii) We considered triangular, square, and hexagonal hypercolumn lattices as possible OD-OP-DP map topologies. The requirement of mutually consistent and symmetrical layout of OD, OP and DP maps has restricted the only possible lattice to be the hexagonal case with vertical OD bands, as shown in Fig.~\ref{fig:hypercol_grid}(C). The triangular lattice is eliminated because it can not satisfy the requirement of having nearest neighboring pinwheels with opposite signs. 

(iii) We demonstrated that there must be an odd number of DP fractures originating from each pinwheel center, in order to cover the full $360^\circ$ of DP in a pair of pinwheels. Furthermore, our arrangement also shows that the DP selectivity vanishes in the middle of DP fractures which connects two PWCs. This explains the petering-out of some DP fractures seen in experiments. The fourfold symmetry of the square lattice is not consistent with an odd number of DP fractures emanating from each PWC, and thus is eliminated.

(iv) We predict that a second type of DP fracture will emanate perpendicularly from the midpoints of fractures that link PWCs, and cross OP contours at approximately 90$^\circ$. Together, these fractures and the ones from Point (iii) account for the experimentally observed relationships between OP contours and DP fractures, including the frequent occurrence of $\sim90^\circ$ changes of direction in the latter.

(v) Our idealized hexagonal lattice has been compared with experimental data and shown to provide a close match. By adjusting the value of $a$ (side length of hexagon), our lattice matches the experimental OD stripe width and pinwheel densities of human, macaque monkey and galago. However, there are mismatches for cat and ferret since these two species have patchy-like OD rather than stripy, and this does not correspond with our approximation of straight and parallel OD bands. Furthermore, because tree shrews and mouse lemurs lack apparent OD pattern, we cannot compare our hexagonal lattice in terms of the OD stripe width for these two animals. Nevertheless, it shows good agreement when comparing to the pinwheel density and mean OP periodicity. In all cases, the mean PWC density and OP periodicity are approximately related by Eqs~(\ref{eq:original_k}) -- (\ref{eq:Lambda_hexa_approx}), which were previously derived for aperiodic cases \citep{ho2021orientation_murinus, Kaschube_PWC_density,keil2012response_to_universality}.

(vi) The OD, OP and DP selectivities within each hypercolumn are discussed in terms of cortical location. Near pinwheel centers the OP and DP selectivity are both low. Moreover, DP selectivity is low near DP fractures, whereas OP selectivity is high in that region, except near PWCs. 

Overall, we have constructed an idealized, maximally symmetric hexagonal lattice with a mutually consistent combined OD-OP-DP map. The hexagonal lattice topology also reproduced previous findings that the layout of OP regions is approximately hexagonal \citep{muir_embedding_2011,paik2011orientation_hexa}. This work can serve as a basis for future studies the effects of adding perturbations to the regular hexagonal lattice to model more realistic, somewhat irregular, feature maps and to derive an analytical operator for combined OD-OP-DP feature detection, and its map to the surface of V1.

\medskip

\acknowledgments
We thank N.~Gale for stimulating discussions. This work was supported by the Australian Research Council under Center of Excellence Grant CE140100007 and the Australian Research Council Laureate Fellowship Grant FL1401000025.


\bibliography{mybib}

\begin{thebibliography}{}

\bibitem[\protect\astroncite{Adams et~al.}{2007}]{adams_complete_2007}
Adams, D.~L., Sincich, L.~C., and Horton, J.~C. (2007).
\newblock Complete pattern of ocular dominance columns in human primary visual
  cortex.
\newblock {\em J Neurosci}, 27(39):10391--10403.

\bibitem[\protect\astroncite{Anderson et~al.}{1988}]{Anderson_cat_Od}
Anderson, P.~A., Olavarria, J., and Van~Sluyters, R.~C. (1988).
\newblock The overall pattern of ocular dominance bands in cat visual cortex.
\newblock {\em J Neurosci}, 8(6):2183--2200.

\bibitem[\protect\astroncite{Balents}{2010}]{balents2010spin}
Balents, L. (2010).
\newblock Spin liquids in frustrated magnets.
\newblock {\em Nature}, 464(7286):199--208.

\bibitem[\protect\astroncite{Bartfeld and
  Grinvald}{1992}]{bartfeld_relationships_1992}
Bartfeld, E. and Grinvald, A. (1992).
\newblock Relationships between orientation-preference pinwheels, cytochrome
  oxidase blobs, and ocular-dominance columns in primate striate cortex.
\newblock {\em Proc Natl Acad Sci USA}, 89:11905--11909.

\bibitem[\protect\astroncite{Blasdel}{1992}]{blasdel_orientation_1992}
Blasdel, G.~G. (1992).
\newblock Orientation selectivity, preference, and continuity in monkey striate
  cortex.
\newblock {\em J Neurosci}, 12:3139--3161.

\bibitem[\protect\astroncite{Blasdel and Salama}{1986}]{blasdel1986voltage}
Blasdel, G.~G. and Salama, G. (1986).
\newblock Voltage-sensitive dyes reveal a modular organization in monkey
  striate cortex.
\newblock {\em Nature}, 321(6070):579.

\bibitem[\protect\astroncite{Bonhoeffer and
  Grinvald}{1991}]{bonhoeffer_iso-orientation_1991}
Bonhoeffer, T. and Grinvald, A. (1991).
\newblock Iso-orientation domains in cat visual cortex are arranged in
  pinwheel-like patterns.
\newblock {\em Nature}, 353:429--31.

\bibitem[\protect\astroncite{Bonhoeffer and
  Grinvald}{1993}]{bonhoeffer_layout_1993}
Bonhoeffer, T. and Grinvald, A. (1993).
\newblock The layout of iso-orientation domains in area 18 of cat visual
  cortex: Optical imaging reveals a pinwheel-like organization.
\newblock {\em J Neurosci}, 13:4157--4180.

\bibitem[\protect\astroncite{Bosking et~al.}{1997}]{bosking_orientation_1997}
Bosking, W.~H., Zhang, Y., Schofield, B., and Fitzpatrick, D. (1997).
\newblock Orientation selectivity and the arrangement of horizontal connections
  in tree shrew striate cortex.
\newblock {\em J Neurosci}, 17(6):2112--2127.

\bibitem[\protect\astroncite{Bressloff}{2002}]{bressloff_PhysRevLet_2002}
Bressloff, P.~C. (2002).
\newblock Bloch waves, periodic feature maps, and cortical pattern formation.
\newblock {\em Phys Rev Lett}, 89(8):088101.

\bibitem[\protect\astroncite{Bressloff}{2003}]{bressloff2003spatially}
Bressloff, P.~C. (2003).
\newblock Spatially periodic modulation of cortical patterns by long-range
  horizontal connections.
\newblock {\em Physica D}, 185(3-4):131--157.

\bibitem[\protect\astroncite{Bressloff and Cowan}{2002}]{bressloff_visual_2002}
Bressloff, P.~C. and Cowan, J.~D. (2002).
\newblock The visual cortex as a crystal.
\newblock {\em Physica D}, 173(3):226--258.

\bibitem[\protect\astroncite{Bressloff and
  Cowan}{2003}]{bressloff_functional_2003}
Bressloff, P.~C. and Cowan, J.~D. (2003).
\newblock The functional geometry of local and horizontal connections in a
  model of {V1}.
\newblock {\em J Physiol Paris}, 97(2):221--236.

\bibitem[\protect\astroncite{Bressloff et~al.}{2001}]{bressloff2001geometric}
Bressloff, P.~C., Cowan, J.~D., Golubitsky, M., Thomas, P.~J., and Wiener,
  M.~C. (2001).
\newblock Geometric visual hallucinations, euclidean symmetry and the
  functional architecture of striate cortex.
\newblock {\em Philos Trans R Soc Lond B Biol Sci}, 356(1407):299--330.

\bibitem[\protect\astroncite{Bressloff et~al.}{2002}]{bressloff_what_2002}
Bressloff, P.~C., Cowan, J.~D., Golubitsky, M., Thomas, P.~J., and Wiener,
  M.~C. (2002).
\newblock What geometric visual hallucinations tell us about the visual cortex.
\newblock {\em Neural Comput}, 14(3):473--491.

\bibitem[\protect\astroncite{Carreira-Perpin{\'a}n
  et~al.}{2004}]{carreira_computation_map_2004}
Carreira-Perpin{\'a}n, M.~{\'A}., Lister, R.~J., and Goodhill, G.~J. (2004).
\newblock {A computational model for the development of multiple maps in
  primary visual cortex}.
\newblock {\em Cerebral Cortex}, 15(8):1222--1233.

\bibitem[\protect\astroncite{Cheng et~al.}{2001}]{cheng2001human_Od}
Cheng, K., Waggoner, R.~A., and Tanaka, K. (2001).
\newblock Human ocular dominance columns as revealed by high-field functional
  magnetic resonance imaging.
\newblock {\em Neuron}, 32(2):359--374.

\bibitem[\protect\astroncite{Crair et~al.}{1997}]{crair1997ocular}
Crair, M.~C., Ruthazer, E.~S., Gillespie, D.~C., and Stryker, M.~P. (1997).
\newblock Ocular dominance peaks at pinwheel center singularities of the
  orientation map in cat visual cortex.
\newblock {\em J Neurophysiol}, 77(6):3381--3385.

\bibitem[\protect\astroncite{Gilbert and
  Wiesel}{1983}]{Gilbert_patchy_connectn}
Gilbert, C. and Wiesel, T. (1983).
\newblock Clustered intrinsic connections in cat visual cortex.
\newblock {\em J Neurosci}, 3(5):1116--1133.

\bibitem[\protect\astroncite{Götz}{1987}]{gotz_d-blob_1987}
Götz, K.~G. (1987).
\newblock Do “d-blob” and “l-blob” hypercolumns tessellate the monkey
  visual cortex?
\newblock {\em Biol Cybern}, 56(2):107--109.

\bibitem[\protect\astroncite{Götz}{1988}]{Gotz_1988}
Götz, K.~G. (1988).
\newblock Cortical templates for the self-organization of orientation-specific
  d- and l-hypercolumns in monkeys and cats.
\newblock {\em Biol Cybern}, 58(4):213--223.

\bibitem[\protect\astroncite{Ho et~al.}{2021}]{ho2021orientation_murinus}
Ho, C. L.~A., Zimmermann, R., Weidinger, J. D.~F., Prsa, M., Schottdorf, M.,
  Merlin, S., Okamoto, T., Ikezoe, K., Pifferi, F., Aujard, F., Angelucci, A.,
  Wolf, F., and Huber, D. (2021).
\newblock Orientation preference maps in microcebus murinus reveal
  size-invariant design principles in primate visual cortex.
\newblock {\em Curr Biol}, 31(4):733--741.

\bibitem[\protect\astroncite{Horton and Adams}{2005}]{Horton_OD}
Horton, J.~C. and Adams, D.~L. (2005).
\newblock The cortical column: A structure without a function.
\newblock {\em Philos Trans R Soc Lond B Biol Sci}, 360(1456):837--62.

\bibitem[\protect\astroncite{Horton and Hocking}{1996}]{horton_intrinsic_1996}
Horton, J.~C. and Hocking, D.~R. (1996).
\newblock Intrinsic variability of ocular dominance column periodicity in
  normal macaque monkeys.
\newblock {\em J Neurosci}, 16(22):7228--7339.

\bibitem[\protect\astroncite{Hubel and Wiesel}{1962a}]{hubel1962receptive}
Hubel, D.~H. and Wiesel, T.~N. (1962a).
\newblock Receptive fields, binocular interaction and functional architecture
  in the cat's visual cortex.
\newblock {\em J Physiol}, 160(1):106--154.

\bibitem[\protect\astroncite{Hubel and
  Wiesel}{1962b}]{Hubel_column_arrangm_1962}
Hubel, D.~H. and Wiesel, T.~N. (1962b).
\newblock Shape and arrangement of columns in cat's striate cortex.
\newblock {\em J Physiol}, 165(3):559--568.

\bibitem[\protect\astroncite{Hubel and Wiesel}{1968}]{hubel1968receptive}
Hubel, D.~H. and Wiesel, T.~N. (1968).
\newblock Receptive fields and functional architecture of monkey striate
  cortex.
\newblock {\em J Physiol}, 195(1):215--243.

\bibitem[\protect\astroncite{Hubel and Wiesel}{1974a}]{hubel_sequence_1974}
Hubel, D.~H. and Wiesel, T.~N. (1974a).
\newblock Sequence regularity and geometry of orientation columns in the monkey
  striate cortex.
\newblock {\em J Comp Neurol}, 158(3):267--293.

\bibitem[\protect\astroncite{Hubel and Wiesel}{1974b}]{hubel1974uniformity}
Hubel, D.~H. and Wiesel, T.~N. (1974b).
\newblock Uniformity of monkey striate cortex: A parallel relationship between
  field size, scatter, and magnification factor.
\newblock {\em J Comp Neurol}, 158(3):295--305.

\bibitem[\protect\astroncite{Hubel and Wiesel}{1977}]{hubel_ferrier_1977}
Hubel, D.~H. and Wiesel, T.~N. (1977).
\newblock Ferrier lecture: Functional architecture of macaque monkey visual
  cortex.
\newblock {\em Proc R Soc Lond B Biol Sci}, 198(1130):1--59.

\bibitem[\protect\astroncite{H{\"u}bener et~al.}{1997}]{hubener1997spatial}
H{\"u}bener, M., Shoham, D., Grinvald, A., and Bonhoeffer, T. (1997).
\newblock Spatial relationships among three columnar systems in cat area 17.
\newblock {\em J Neurosci}, 17(23):9270--9284.

\bibitem[\protect\astroncite{Hübener and Bonhoeffer}{2002}]{HUBENER2002131}
Hübener, M. and Bonhoeffer, T. (2002).
\newblock Optical imaging of functional architecture in cat primary visual
  cortex.
\newblock In Bertram, R.~P. and Peters, A., editors, {\em The Cat Primary
  Visual Cortex}, pages 131--167. Academic Press, San Diego.

\bibitem[\protect\astroncite{Kaschube et~al.}{2010}]{Kaschube_PWC_density}
Kaschube, M., Schnabel, M., L{\"o}wel, S., Coppola, D.~M., White, L.~E., and
  Wolf, F. (2010).
\newblock Universality in the evolution of orientation columns in the visual
  cortex.
\newblock {\em Science}, 330(6007):1113--1116.

\bibitem[\protect\astroncite{Keil
  et~al.}{2012}]{keil2012response_to_universality}
Keil, W., Kaschube, M., Schnabel, M., Kisvarday, Z.~F., L{\"o}wel, S., Coppola,
  D.~M., White, L.~E., and Wolf, F. (2012).
\newblock Response to comment on ``universality in the evolution of orientation
  columns in the visual cortex''.
\newblock {\em Science}, 336(6080):413--413.

\bibitem[\protect\astroncite{Kisv{\'a}rday
  et~al.}{2001}]{kisvarday2001calculatingDP}
Kisv{\'a}rday, Z.~F., Buz{\'a}s, P., and Eysel, U.~T. (2001).
\newblock Calculating direction maps from intrinsic signals revealed by optical
  imaging.
\newblock {\em Cerebral Cortex}, 11(7):636--647.

\bibitem[\protect\astroncite{Law et~al.}{1988}]{law1988organization_ferret}
Law, M.~I., Zahs, K.~R., and Stryker, M.~P. (1988).
\newblock Organization of primary visual cortex (area 17) in the ferret.
\newblock {\em J Comp Neurol}, 278(2):157--180.

\bibitem[\protect\astroncite{Levay et~al.}{1978}]{levay_cat_Od}
Levay, S., Stryker, M.~P., and Shatz, C.~J. (1978).
\newblock Ocular dominance columns and their development in layer {IV} of the
  cat's visual cortex: A quantitative study.
\newblock {\em J Comp Neurol}, 179(1):223--244.

\bibitem[\protect\astroncite{Liu and Robinson}{2021}]{liu2021analytic}
Liu, X. and Robinson, P.~A. (2021).
\newblock Analytic model for feature maps in the primary visual cortex.
\newblock {\em arXiv preprint arXiv:2103.09954}.

\bibitem[\protect\astroncite{Maldonado et~al.}{1997}]{maldonado1997orientation}
Maldonado, P.~E., G{\"o}decke, I., Gray, C.~M., and Bonhoeffer, T. (1997).
\newblock Orientation selectivity in pinwheel centers in cat striate cortex.
\newblock {\em Science}, 276(5318):1551--1555.

\bibitem[\protect\astroncite{Miikkulainen
  et~al.}{2005}]{Miikkulainen_visual_maps}
Miikkulainen, R., Bednar, J.~A., Choe, Y., and Sirosh, J. (2005).
\newblock {\em Computational Maps in the Visual Cortex}.
\newblock Springer-Verlag, New York.

\bibitem[\protect\astroncite{Muir et~al.}{2011}]{muir_embedding_2011}
Muir, D.~R., Costa, D., M., N., Girardin, C.~C., Naaman, S., Omer, D.~B.,
  Ruesch, E., Grinvald, A., and Douglas, R.~J. (2011).
\newblock Embedding of cortical representations by the superficial patch
  system.
\newblock {\em Cereb Cortex}, 21.

\bibitem[\protect\astroncite{Müller et~al.}{2000}]{muller_analysis_2000}
Müller, T.~M., Stetter, M., Hübener, M., Sengpiel, F., Bonhoeffer, T.,
  Gödecke, I., Chapman, B., Löwel, S., and Obermayer, K. (2000).
\newblock An analysis of orientation and ocular dominance patterns in the
  visual cortex of cats and ferrets.
\newblock {\em Neural Comput}, 12(11):2573--2595.

\bibitem[\protect\astroncite{Nauhaus et~al.}{2008}]{nauhaus2008neuronal}
Nauhaus, I., Benucci, A., Carandini, M., and Ringach, D.~L. (2008).
\newblock Neuronal selectivity and local map structure in visual cortex.
\newblock {\em Neuron}, 57(5):673--679.

\bibitem[\protect\astroncite{Obermayer and
  Blasdel}{1993}]{obermayer_geometry_1993}
Obermayer, K. and Blasdel, G.~G. (1993).
\newblock Geometry of orientation and ocular dominance columns in monkey
  striate cortex.
\newblock {\em J Neurosci}, 13(10):4114--4129.

\bibitem[\protect\astroncite{Obermayer and
  Blasdel}{1997}]{obermayer1997singularities}
Obermayer, K. and Blasdel, G.~G. (1997).
\newblock Singularities in primate orientation maps.
\newblock {\em Neural Computation}, 9(3):555--575.

\bibitem[\protect\astroncite{Obermayer
  et~al.}{1992}]{obermayer_statistical-mechanical_1992}
Obermayer, K., Blasdel, G.~G., and Schulten, K. (1992).
\newblock Statistical-mechanical analysis of self-organization and pattern
  formation during the development of visual maps.
\newblock {\em Phys Rev A}, 45(10):7568--7589.

\bibitem[\protect\astroncite{Ohki et~al.}{2006}]{ohki2006highlyorderpinwheel}
Ohki, K., Chung, S., Kara, P., H{\"u}bener, M., Bonhoeffer, T., and Reid, R.~C.
  (2006).
\newblock Highly ordered arrangement of single neurons in orientation
  pinwheels.
\newblock {\em Nature}, 442(7105):925--928.

\bibitem[\protect\astroncite{Paik and Ringach}{2011}]{paik2011orientation_hexa}
Paik, S. and Ringach, D.~L. (2011).
\newblock Retinal origin of orientation maps in visual cortex.
\newblock {\em Nat Neurosci}, 14(7):919--925.

\bibitem[\protect\astroncite{Shatz and Stryker}{1978}]{shatz1978ocular}
Shatz, C.~J. and Stryker, M.~P. (1978).
\newblock Ocular dominance in layer {IV} of the cat's visual cortex and the
  effects of monocular deprivation.
\newblock {\em The Journal of physiology}, 281(1):267--283.

\bibitem[\protect\astroncite{Shmuel and Grinvald}{1996}]{shmuel1996functional}
Shmuel, A. and Grinvald, A. (1996).
\newblock Functional organization for direction of motion and its relationship
  to orientation maps in cat area 18.
\newblock {\em J Neurophysiol}, 16(21):6945--6964.

\bibitem[\protect\astroncite{Swindale}{1996}]{swindale_review_1996}
Swindale, N.~V. (1996).
\newblock The development of topography in the visual cortex: A review of
  models.
\newblock {\em Network}, 7(1):161--247.

\bibitem[\protect\astroncite{Swindale
  et~al.}{2003}]{swindale2003spatialpattern}
Swindale, N.~V., Grinvald, A., and Shmuel, A. (2003).
\newblock The spatial pattern of response magnitude and selectivity for
  orientation and direction in cat visual cortex.
\newblock {\em Cereb Cortex}, 13(3):225--238.

\bibitem[\protect\astroncite{Swindale et~al.}{1987}]{swindale1987surface}
Swindale, N.~V., Matsubara, J.~A., and Cynader, M.~S. (1987).
\newblock Surface organization of orientation and direction selectivity in cat
  area 18.
\newblock {\em J Neurophysiol}, 7(5):1414--1427.

\bibitem[\protect\astroncite{Veltz et~al.}{2015}]{Veltz_2015}
Veltz, R., Chossat, P., and Faugeras, O. (2015).
\newblock On the effects on cortical spontaneous activity of the symmetries of
  the network of pinwheels in visual area {V1}.
\newblock {\em J Math Neurosci}, 5(1):11.

\bibitem[\protect\astroncite{Wang et~al.}{2014}]{wang2014motion}
Wang, H.~X., Merriam, E.~P., Freeman, J., and Heeger, D.~J. (2014).
\newblock Motion direction biases and decoding in human visual cortex.
\newblock {\em J Neurophysiol}, 34(37):12601--12615.

\bibitem[\protect\astroncite{Wannier}{1950}]{wannier1950antiferromagnetism}
Wannier, G.~H. (1950).
\newblock Antiferromagnetism. the triangular ising net.
\newblock {\em Phys Rev}, 79(2):357.

\bibitem[\protect\astroncite{Weliky et~al.}{1996}]{weliky1996DPsystematic}
Weliky, M., Bosking, W.~H., and Fitzpatrick, D. (1996).
\newblock A systematic map of direction preference in primary visual cortex.
\newblock {\em Nature}, 379(6567):725--728.

\bibitem[\protect\astroncite{Xu et~al.}{2005}]{xu2005functional_map_galago}
Xu, X., Bosking, W.~H., White, L.~E., Fitzpatrick, D., and Casagrande, V.~A.
  (2005).
\newblock Functional organization of visual cortex in the prosimian bush baby
  revealed by optical imaging of intrinsic signals.
\newblock {\em J Neurophysiol}, 94(4):2748--2762.

\bibitem[\protect\astroncite{Yacoub et~al.}{2008}]{yacoub_human_op_map}
Yacoub, E., Harel, N., and Uğurbil, K. (2008).
\newblock High-field {fMRI} unveils orientation columns in humans.
\newblock {\em Proc Natl Acad Sci USA}, 105(30):10607--10612.

\end{thebibliography}

\end{document}